\documentclass[preprintnumbers, floatfix,letterpaper, nofootinbib, twocolumn,aps,prd,epsfig]{revtex4-1}
\usepackage{bm,graphicx,dcolumn,epstopdf,epsf, latexsym,mathbbol, amssymb,amsmath,color,slashed, mathrsfs,mathcomp,simplewick}
\usepackage[center]{subfigure}
\usepackage{multirow}
\usepackage{makecell}
\usepackage[colorlinks,linkcolor=blue,citecolor=blue,urlcolor=blue]{hyperref}
\usepackage[center]{subfigure}
\usepackage{relsize,exscale}

\begin{document}
\allowdisplaybreaks
 \newcommand{\bq}{\begin{equation}}
 \newcommand{\eq}{\end{equation}}
 \newcommand{\bqn}{\begin{eqnarray}}
 \newcommand{\eqn}{\end{eqnarray}}
 \newcommand{\ban}{\begin{align}}
 \newcommand{\ean}{\end{align}}
  \newcommand{\nb}{\nonumber}
 \newcommand{\lb}{\label}
 \newcommand{\f}{\frac}
 \newcommand{\p}{\partial}
\newcommand{\PRL}{Phys. Rev. Lett.}
\newcommand{\PLB}{Phys. Lett. B}
\newcommand{\PRD}{Phys. Rev. D}
\newcommand{\CQG}{Class. Quantum Grav.}
\newcommand{\JCAP}{J. Cosmol. Astropart. Phys.}
\newcommand{\JHEP}{J. High. Energy. Phys.}
\newcommand{\NPB}{Nucl. Phys. B}
\newcommand{\Doi}{https://doi.org}
\newcommand{\red}{\textcolor{black}}
\newcommand{\magenta}{\textcolor{black}}
\newcommand{\magggg}{\textcolor{black}}
\newcommand{\gre}{\textcolor{black}}
\title{ Imprints of dark matter on gravitational ringing of supermassive black holes}


\author{Chao Zhang ${}^{a, b, c}$}
\email{chao123@zjut.edu.cn}

\author{Tao Zhu${}^{a, b}$}
\email{corresponding author:  zhut05@zjut.edu.cn}

\author{Xiongjun Fang${}^{d}$}
\email{fangxj@hunnu.edu.cn}

\author{Anzhong Wang${}^{e}$}
 \email{anzhong$\_$wang@baylor.edu}

\affiliation{
${}^{a}$ Institute for theoretical physics and cosmology, Zhejiang University of Technology, Hangzhou, 310032, China\\
${}^{b}$ United Center for Gravitational Wave Physics (UCGWP), Zhejiang University of Technology, Hangzhou, 310032, China\\
${}^{c}$ College of Information Engineering, Zhejiang University of Technology, Hangzhou, 310032, China \\
${}^{d}$ Department of Physics, Key Laboratory of Low Dimensional Quantum Structures and Quantum Control of Ministry of Education, and Synergetic Innovation Center for Quantum Effects and Applications, Hunan Normal University, Changsha, Hunan 410081, China\\
${}^{e}$ GCAP-CASPER, Physics Department, Baylor University, Waco, TX 76798-7316, USA}

\date{\today}

\begin{abstract}

Gravitational waves emitted from the gravitational ringing of supermassive black holes are important targets to test general relativity and probe the matter environment surrounding such black holes. The main components of the ringing waveform are black hole quasi-normal modes. In this paper, we study the effects of the dark matter halos with three different density profiles on the gravitational polar (even-parity) perturbations of a supermassive black hole. For this purpose, we first consider modified Schwarzschild spacetime with three different dark matter profiles and derive the equation of motion of the polar perturbations of the supermassive black hole. It is shown that by ignoring the dark matter perturbations, a Zerilli-like master equation with a modified potential for the polar perturbation can be obtained explicitly. Then we calculate the complex frequencies of the quasi-normal modes of the supermassive black hole in the dark matter halos. The corresponding gravitational wave spectra with the effects of the dark matter halos and their detectability have also been discussed.

\end{abstract}

\maketitle

\section{Introduction}
\renewcommand{\theequation}{1.\arabic{equation}} \setcounter{equation}{0}

Black holes (BHs) are one of the most mysterious phenomena in the universe. The existence of BHs provides us a perfect way to test gravitational effects under extremely strong gravitational fields, such as the formation of gigantic jets of particles and disruption of neighboring stars. On the other hand, from the theoretical point of view, BHs are also excellent labs to test modified theories of gravity that are different from general relativity (GR) (see, e.g., \cite{test_GR1, test_GR2, test_GR3, Xiang2019, Chao2020, Chao2020b, Berti2016}).

It is widely believed that the central region of many galaxies contain supermassive BHs \cite{Melia2001, Genzel2010}. That's one of the reasons that the detection of the shadow of the M87 central BH  with the Event Horizon Telescope (EHT) \cite{EHTL1, EHTL2, EHTL3, EHTL4, EHTL5, EHTL6, EHTL7, EHTL8} drew a lot of attentions. Interestingly, this shadow image is in good agreement with the prediction of the spacetime geometry of a BH described by the Kerr metric \cite{Kerr1963}. Nonetheless, since it is believed that up to $ 90\%$ of the matter in a host galaxy is made up by dark matter (DM) \cite{Kimet2020}, it is natural to expect that the DM halo surrounding a central BH will bring small deviations to the Kerr metric. That leads us to the study of the influence of DM halos in galaxies.

In fact, a lot of research has been done on DM and proposed various black hole models immersed in the DM halos \cite{Zhaoyi2020, Kimet2019, kimet_shadow, xu_JCAP, Xu:2021dkv}. {Of particular interest is the GWs emitted from the supermassive black hole during the ringdown phase of a binary supermassive black hole merger or a extreme mass-ratio inspiring. Such GWs can be described by linear metric perturbations about the black hole spacetime with dark matter halos  \cite{Poisson2005}. In fact, in the background spacetime which is static and spherically symmetric, the metric perturbations can decouple into two independent parts, the polar one and the axial one. As we discuss later, the polar perturbation is equivalent to the even-parity perturbation, which corresponds to the parity of $(-1)^{l+1}$, with $l$ being an index inherited from spherical harmonics \cite{Cardoso2001}. In contrast, the axial perturbation corresponds to the parity of  $(-1)^{l}$.}

{In principle, we can't fully describe the physics of the corresponding GW solely by the axial sector or by the polar sector, at least not {\it a priori}. It is their combination form the full waveform of the corresponding GWs \cite{Poisson2005}. For this reason, in order to extract the full GW waveform from the gravitational ringings, one has to consider both the polar and axial perturbations.}
In \cite{Chao2021}, we focused on the effects of DM halos on gravitational waves (GWs) emitted from  the gravitational axial perturbations of the central BH located in a galaxy. Note that in \cite{Liu:2021xfb}, the effects of two specific dark matter profiles on the gravitational ringing of axial perturbations have also been studied. By describing these supermassive BHs with Schwarzschild-like spacetimes under different DM models, we investigated the gravitational perturbations and calculated the corresponding quasi-normal modes (QNMs) \cite{Chao2021}. What's more, the effects of some model-dependent parameters were also studied. This paper is a successor of our previous work. Since in \cite{Chao2021} only the axial-perturbation sector has been investigated, here we shall move to the polar perturbations by using the technique developed recently in \cite{Wentao2021} for constructing master equation of polar perturbations for spherical symmetric BHs.

It is also the enthusiasm on GWs pushes us to investigate the physics within that ever since its first detection from the coalescence of two massive BHs by advanced LIGO, which marked the beginning of a new era, the GW astronomy \cite{Ref1}. Following this observation, about 90 GW events have been identified by the LIGO/Virgo/KAGRA scientific collaboration (see, e.g., \cite{GWs,GWs19a,GWs19b, GWsO3b}). In the future, more ground- and space-based GW detectors will be constructed \cite{Moore2015, Aso2013, Liu2020, Taiji2, Gong:2021gvw}, which will enable us to probe signals with a wider frequency band. This actually triggered the interests on the QNMs from GWs, including those from the late-merger and ringdown stages of a coalescence \cite{Berti18} as well as those from supermassive BHs, e.g, central BHs in a galaxy. The detection of QNMs from the ringdown stage will provide a unique way to probe the matter environment surrounding a BH. For example, it is shown that the shift from GR on the QNMs due to the surrounding ultralight bosons might be detectable in the future by the observational data from LISA-like missions \cite{Chung:2021roh, Brito2017}. The environmental effects such as dark matter halos on the BH ringdown emission and their implications on the GW detections have been widely studied in \cite{Paolo}. The effects of the DM or matter distributing around BHs on the QNMs have also been considered in \cite{Cardoso:2021wlq, Liu:2021xfb, Konoplya2021, Chao2021}.

From the theoretical point of view, QNMs are eigenmodes of dissipative systems. The information contained in QNMs provide the keys in revealing whether BHs are ubiquitous in our universe, and more important whether GR is the correct theory to describe the event even in the strong field regime. Readers may find more details in \cite{Berti2009}. Basically, the QNM frequency $\omega$ contains two parts, the real part and the imaginary part. Its real part gives the frequency of vibration while its imaginary part provides the damping time. In other words, the frequency we are going to calculate is a complex number (although it could be purely imaginary in certain circumstances).

According to GR, to extract the physics from QNMs, at least two QNM signals are needed. This will require the signal-to-noise ratio (SNR) to be of the order 100 \cite{Chao2021}. Although such high SNRs are not achievable right now, it has been shown that they may be achievable once the advanced LIGO and Virgo reach their designed sensitivities. In any case, it is certain that they will be detected by the ground-based third-generation detectors, such as Cosmic Explorer or the Einstein Telescope, as well as the space-based detectors, including LISA, TianQin \cite{Shi2019}, Taiji \cite{Taiji2}, and DECIGO \cite{Moore2015}.

As mentioned above, QNMs could be calculated under the polar or axial perturbations. In general, we expect deviations between these two cases. However, for the Schwarzschild case, the QNMs capture an interesting property that is referred as isospectrality \cite{Berti2009}, which is a portmanteau meaning that the spectra from the polar and axial perturbations are identical. Note that this could be proven analytically \cite{Chand83}. Inspiring by this, we shall try to test if isospectrality is preserved for our problem. Specially, we want to test if we can break isospectrality for certain scenarios under our considerations. This is an important aspect to label the difference between the Schwarzschild and non-Schwarzschild spacetimes.

For this purpose, we first consider modified Schwarzschild BHs with three different DM profiles and derive the master equation of the polar (even-parity) perturbations for the calculations of QNMs. It is shown that by ignoring the DM perturbations,  a Zerilli-like master equation with a modified potential for the polar perturbation can be obtained explicitly. Then we calculate the frequencies of the QNMs of the Schwarzschild-like BH in the DM halos. The corresponding GW spectra with the effects of the DM halos have also been discussed. Notice that similar scenarios for the axial (odd-parity) case have been considered in \cite{Chao2021}. The phenomena we have observed there will be treated as comparisons to some of the results in this current paper. And of course, we will also study the possibility of breaking the isospectrality.

Here, we shall consider the QNMs of the polar metric perturbaions of a Schwarzschild-like BH sorrounded by the DM halos. Several different background metrics are investigated by considering three different DM profiles. These metrics for different DM halo profiles can be found in \cite{Kimet2019, xu_JCAP}. Note that recently the metric of a BH immersed in DM spike has also been derived \cite{Xu:2021dkv}. In addition, the Sgr $\text{A}^\ast$ BH (located in the center of Milky Way galaxy) and the M87 galactic central BH are what we focus on (Some of their extensions will also be studied). In other words, the structure we will consider is a BH located at the center of a galaxy. By comparing the resultant QNMs with their counterparts for the Schwarzschild case and the axial-perturbations, we will see the influence of a DM halo on QNMs as well as GWs of these central supermassive BHs.

The rest of this paper is organized as follows: Sec. II shows some basic information of the three density profiles of DM halos that we are going to investigate for the calculations of QNMs. After that, in Sec. III we show briefly how to derive a Zerilli-like master equation from the polar perturbation and Einstein's field equations. Sec. IV contains three parts. In the first part we present some resultant QNMs. Some concluding remarks will be given by comparing them with their counterparts in the Schwarzschild case with axial-perturbations. For the second part, we focus ourselves on the $l=2$ case and will test the effects of model-dependent constants. In the last part we discuss the isospectrality of our problem. After that, in Sec. V we investigate in detail how QNMs deviate from that of the Schwarzschild case  by changing the model-dependent parameters, which reveals the detectability of these deviations. Finally, Sec. VI provides our main conclusions as well as some outlooks to the future work.

Through out the paper, we shall adopt the unit system so that $c=G_N=1$, where $c$ is the speed of light while $G_N$ stands for the  gravitational constant observed on Earth. In this way, we still have one degree of freedom to choose the unit for length. This will be done later by setting $r_{MH}^{\text{Sch}}=1$, where $r_{MH}^{\text{Sch}}$ is the radius of the metric horizon (MH) for the Schwarzshild BH and we have $ r_{MH}^{\text{Sch}} \equiv 2 G_N M/c^2$, with $M$ being the mass of the BH that we are focusing on. {In this paper, all the Greek indices run from 0 to 3. Other usage of indices will be indicated explicitly when it is necessary.}

\section{Black hole solutions in dark matter halo}
\renewcommand{\theequation}{2.\arabic{equation}} \setcounter{equation}{0}

Normally, the black hole spacetimes are not clean, and are affected by the surrounding matters. In this section, we consider the spherically symmetric static black hole solutions with several different DM halo profiles. A Schwarzshild  BH in the DM halo could be described by the metric (as assumed in \cite{xu_JCAP, Xu:2021dkv}),
\bqn
\lb{metric_dark}
ds^2 & = & - G(r) dt^2 + \frac{1}{F(r)} dr^2 + r^2 d\theta^2 + r^2 \sin^2\theta d\varphi^2, \nb\\
\eqn
where $G(r)$ and $F(r)$ denote the functions that describe the effects of the DM halos and BH on the metric. As mentioned in \cite{xu_JCAP, Xu:2021dkv}, if one ignores the high-order effects of potential of dark matter (they are supposed to be small and very complicated to model) and assume the deformed metrics satisfy Newtonian approximation, then one can set $F(r)=G(r)$.
Without special alerts, in the following we shall avoid mentioning $F(r)$ and just assume $G(r)=F(r)$. With such setup, then the deformed metrics with different dark matter halos can be constructed by analyzing the Einstein's field equations \footnote{{ For certain scenarios, other types of metrics and approaches to obtain their solutions are also discussed in the literature. See, e.g., \cite{Cardoso:2021wlq, Cardoso2022, Konoplya2022, Kimet2022}.} } \cite{xu_JCAP, Xu:2021dkv}. These deformed metrics satisfy the Einstein's field equations and reduce to the Schwarzschild solution, $G(r)=1- 2M/r$ \cite{CarrollB}, when the DM is absent. For different profiles of the DM halo, the function $G(r)$ is different \cite{xu_JCAP}. In the following, we are going to present the function $G(r)$ for each profile individually [For  readers to see more details about solving for $G(r)$, e.g., the explicit form of the stress-energy tensor in Einstein's field equations, we suggest \cite{xu_JCAP} as a reference].

\subsection{URC profile}

In the universal rotation curve (URC) profile of the DM halo, the distribution of the DM is described by \cite{URC1} (see also \cite{halo_review} for a review)
\bqn
\rho(r) = \frac{\rho_0 r_0^3}{(r+r_0)(r^2+r_0^2)},
\eqn
where $\rho_0$ is the central density and $r_0$ is the characteristic radius of the DM halo. According to the observations on the M87 galaxy, the best fit values for the parameters of the URC profile are $\rho_0 = 6.9\times 10^6 \text{M}_{\odot}/{\rm kpc}^{3}$ and $r_0 = 91.2\; {\rm kpc}$ \cite{Salucci_M87}. While in the Milky Way galaxy, we have $\rho_0 = 5.2 \times 10^7 \text{M}_{\odot}/{\rm kpc}^{3}$ and $r_0 = 7.8\; {\rm kpc}$ \cite{dark_matter}. With this halo profile, the function $G(r)$ in the metric (\ref{metric_dark}) is given by \cite{Kimet2019, kimet_shadow}  \footnote{Note that, Eq.~\eqref{URC_Gr} is different from its counterpart in \cite{Kimet2019}, viz., Eq.~(18), up to a factor $e^{-2 \pi ^2 \rho_0 r_0^2}$. Such a factor is added in Eq.~\eqref{URC_Gr} for the function $G(r)$ to be normalized at the spatial infinity.}
\bqn
\lb{URC_Gr}
G(r) &=& e^{-2 \pi ^2 \rho_0 r_0^2} \left(1+\frac{r^2}{r_0^2}\right)^{-\frac{2 \,\rho_0 r_0^3 \pi}{r}(1-\frac{r}{r_0})} \nb\\
&& \times \left(1+\frac{r}{r_0} \right)^{-\frac{4\, \rho_0 r_0^3  \pi}{r}(1+\frac{r}{r_0})} \nb\\
&& \times \exp\left[\frac{4\, \rho_0 r_0^3 \pi \arctan(\frac{r}{r_0})(1+\frac{r}{r_0})}{r}   \right]- \dfrac{2M}{r}.\nb\\
\eqn
Here $M = 6.5 \times 10^9~\text{M}_{\odot}$ for the M87 central BH and $M = 4.3 \times 10^6~\text{M}_{\odot}$ for the Sgr $\text{A}^\ast$ BH.

\subsection{The CDM halo with NFW profile}

The cold dark matter (CDM) halo with Navarro-Frenk-White (NFW) profile is obtained by $N$-body simulations, which has a universal spherically averaged density profile \cite{NFW, Kimet2019},
\bqn
\rho(r) = \frac{\rho_0}{(r/r_0)(1+r/r_0)^2},
\eqn
where $\rho_0$ is the density of the universe at the moment when the halo collapsed and $r_0$ is the characteristic radius.  According to the observations on Milky Way galaxy \cite{dark_matter}, the best fit values for the parameters $\rho_0$ and $r_0$ for NFW profile are $\rho_0 = 5.23 \times 10^7 \text{M}_{\odot}/{\rm kpc}^{3}$ and $r_0=8.1 \; {\rm kpc}$. On the other hand, for M87 galaxy we will have  $\rho_0 = 0.008 \times 10^{7.5}~\text{M}_{\odot}/ \text{kpc}^3$ (see \cite{Oldham2016}) and ${r_0} = 130~\text{kpc}$ \cite{Kimet2019}. With this halo profile, the function $G(r)$ in the metric (\ref{metric_dark}) is given by \cite{xu_JCAP}
\bqn
\lb{Gr2}
G(r) = \left(1+\frac{r}{{r_0}}\right)^{-\frac{8 \pi G_N \rho_0 r_0^3}{c^2 r}} - \frac{2 G_N M}{c^2 r}.
\eqn
Here $M = 4.3 \times 10^6~\text{M}_{\odot}$ is the mass of Sgr $\text{A}^\ast$ BH and $M = 6.5 \times 10^9~\text{M}_{\odot}$ is the mass of M87 central BH.

\subsection{The SFDM model}

For the Scalar Field Dark Matter (SFDM) model \cite{Xian2018, xu_JCAP}, the energy density profile for DM halo is given by
\bqn
\lb{rho3}
\rho(r) = \frac{\rho_0 \sin(\pi r/r_0)}{\pi r/r_0},
\eqn
where $\rho_0$ is the central density and $r_0$ is the radius at which the pressure and density are zero. In Milky Way galaxy, we have $\rho_0 = 3.43\times 10^7 \text{M}_{\odot}/{\rm kpc}^{3}$ and $r_0 = 15.7 \; {\rm kpc}$ \cite{Xian2018}. With this halo profile, the function $G(r)$ in the metric (\ref{metric_dark}) is given by
\bqn
\lb{Gr3}
G(r) = \exp\left[-\frac{8 G_N \rho_0 r_0^2}{\pi} \frac{\sin(\pi r/r_0)}{\pi r/r_0}\right] - \frac{2 G_N M}{c^2 r}. \nb\\
\eqn
Here $M = 4.3 \times 10^6~\text{M}_{\odot}$ is the mass of Sgr $\text{A}^\ast$ BH.

 \section{Zerilli-like equation for polar metric perturbations}
\renewcommand{\theequation}{3.\arabic{equation}} \setcounter{equation}{0}

In this section, we consider the linear gravitational perturbations $h_{\mu\nu}$ around Schwarzshild-like solutions. Let us first start with a general form of a spherically symmetric spacetime, given by \cite{Wentao2021}
\bqn
\lb{backg}
d s^2 &=& -e^{2 \Phi(r)} dt^2 + e^{2 \Lambda(r)} d r^2+ r^2 d\Omega^2,
\eqn
where
\bqn
\lb{dOmega}
d \Omega^2 &=& d \theta^2+\sin^2\theta d \phi^2.
\eqn
Of course, for our case [cf. \eqref{metric_dark}], we have $\Phi=\ln(G)/2$ and $\Lambda=-\ln(G)/2$ (recall that we hve assmued $F=G$). For metric perturbations, let us start by describing the geometry of a linearly perturbed spherically symmetric background $\bar g_{\mu\nu}$,
\bqn
\lb{pertb1}
{g_{\mu\nu} = \bar g_{\mu\nu} + h_{\mu\nu}, }
\eqn
where
\bqn
\lb{backg2}
\bar g_{\mu\nu} = {\rm diag}\left(-e^{2 \Phi(r)}, e^{2 \Lambda(r)}, r^2,r^2 \sin^2\theta\right),~~
\eqn
with $h_{\mu\nu}$ denoting the linear perturbations of the background metric $\bar g_{\mu\nu}$. In general, the perturbation $h_{\mu\nu}$ can be split into pieces that transforming as scalars, vectors, and tensors with respect to the symmetry of the spacetime. However, in two-dimension maximally symmetric space $S^2$, it can be shown that the tensor perturbations with transverse-traceless condition are identically zero \cite{cai_generalized_2013, takahashi_hawking_2010, takahashi_master_2010}. Thus, the metric perturbations can be split as scalar and vector perturbations, i.e., $h_{\mu\nu} =h^{\rm S}_{\mu \nu} + h^{\rm V}_{\mu\nu}$. Here we note that the scalar perturbation is also called polar-type perturbation (or even-parity perturbation) while the vector one is called axial perturbation (or odd-parity perturbation) \cite{Berti2009}. In this paper, for simplicity, we only focus on the polar perturbations of the Schwarzshild-like solutions with different dark matter halos (The axial-perturbation case has already been  studied in \cite{Chao2021, Liu:2021xfb}).

We parameterize the polar perturbations in the form of \cite{Thomp2017}
\begin{widetext}
	\bqn
	\lb{hab}
h_{\mu \nu} &=& \sum_{l=0}^{\infty} \sum_{m=-l}^{l}
	\begin{pmatrix}
		 A_{lm}	& -D_{lm} & -r B_{lm} \partial_\theta & -r B_{lm} \partial_\varphi\\
		 -D_{lm}	&  K_{lm} & r H_{lm} \partial_\theta & r H_{lm} \partial_\varphi\\
		sym	& sym &  r^2 \left[E_{lm}+F_{lm} \left(\partial^2_\theta+L/2\right)\right] &  sym \\
		sym	& sym & r^2 F_{lm} \left(\partial_\theta \partial_\varphi-\cot \theta \partial_\varphi \right) &  r^2 \sin^2 \theta \left[E_{lm} - F_{lm} \left(\partial^2_\theta+L/2\right)\right]
	\end{pmatrix}
	Y_{l m}(\theta, \varphi)  \epsilon, 	\nb\\
	\eqn
\end{widetext}
where $A_{lm}$, $D_{lm}$, $H_{lm}$, $K_{lm}$, $B_{lm}$, $E_{lm}$ and $F_{lm}$ are functions of $t$ and $r$. $Y_{l m}(\theta, \varphi)$ stands for the spherical harmonics \cite{Zettilib} {and $l$ as well as $m$ in the index are integers}. In addition, we have defined $L \equiv l(l+1)$. Here, $\epsilon$ is a real number and $|\epsilon| \ll 1$.

From now on, we will set  $m=0$ in \eqref{hab} so that $\partial_\varphi Y_{lm}(\theta, \varphi)=0$, as now the background has the spherical symmetry, and the corresponding linear perturbations do not depend on $m$   \cite{Regge57,Thomp2017}. In addition, by adopting the RW gauge \cite{Thomp2017}, we will set $B_{lm}=F_{lm}=H_{lm}=0$.

By following \cite{Wentao2021} and using Einstein's field equations \cite{CarrollB}, for the vacuum case we obtain \footnote{For simplicity, here we ignore the perturbation of the DM since its effects are expected to be negligible in comparing to the effects of DM from the modified background geometry \cite{Chao2021}.}
\bqn
\lb{PDE1}
0 &=& { \left[ \eta_1 \frac{\partial^2}{\partial r^2} + \eta_2 \frac{\partial}{\partial r} +\left(- \frac{\partial^2}{\partial t^2} +\eta_3  \right) \right] Z^{\text{II}} (t, r), }
\eqn
with
\bqn
\lb{eta123}
\eta_1 &\equiv& \frac{N^{\text{II}} \left(\eta^{\text{II}}\right)^2 \sigma^{\text{II}} \left(\tau^{\text{II}}\right)^2}{e^{2 (\Lambda-\Phi)} N_T^{\text{II}}}, \nb\\
\eta_2 &\equiv& \frac{N_R^{\text{II}} \left(\eta^{\text{II}}\right)^2 \tau^{\text{II}} }{2 r e^{2 (\Lambda-\Phi)} N_T^{\text{II}}}, \nb\\
\eta_3 &\equiv& - \frac{N_Z^{\text{II}} \left(\eta^{\text{II}}\right)^2  }{2 r^2 e^{2 (\Lambda-\Phi)} N_T^{\text{II}}},
\eqn
where $N^{\text{II}}$, $N_T^{\text{II}}$, $N_R^{\text{II}}$, $N_Z^{\text{II}}$, $\sigma^{\text{II}}$, $\eta^{\text{II}}$ and $\tau^{\text{II}}$ are functions of $\Phi(r)$, $\Lambda(r)$, $r$ as well as $L$ and are defined in \cite{Wentao2021}. Notice that $ Z^{\text{II}}(t, r)$ is a gauge invariant constructed by $A_{lm}$, $D_{lm}$, $K_{lm}$ and  $E_{lm}$ \footnote{ { Notice that, once  $ Z^{\text{II}}(t, r)$ has been achieved, the non-vanishing components $A_{lm}$, $D_{lm}$, $K_{lm}$ and  $E_{lm}$ can be obtained.}}. In addition, we have dropped the $lm$ in its subscript for simplicity. {We have used the fact that $\Lambda=-\Phi$ (for our case) in simplifying the above expression.} After that, by modifying \eqref{PDE1}, we obtain the master equation
\bqn
\lb{master1}
\frac{d^2 \Psi(t, x)}{d x^2}-\left[\frac{d^2 }{d t^2} + V_{\rm eff}(r)\right] \Psi(t, x)  &=& 0 ,
\eqn
where
\bqn
\lb{Psi}
\Psi &\equiv& \eta_1^{-1/4}\exp \left( \frac{1}{2}  {  \int {\frac{\eta_2}{\eta_1} dr}} \right) Z^{\text{II}},~~~~\\ \nb\\
\lb{rast}
\frac{d r}{d x} &=& \eta_1^{1/2},
\eqn
and the effective potential is given by
\begin{widetext}
\bqn
\lb{Veff}
V_{\rm eff} &\equiv& { - \left\{\frac{1}{4} \left[2 \eta_1^{1/2} \left(\frac{d \sqrt{\eta_1}}{dr}-\frac{\eta_2}{\sqrt{\eta_1}} \right)^{\prime}-\left(\frac{d \sqrt{\eta_1}}{dr}-\frac{\eta_2}{\sqrt{\eta_1}} \right)^2\right]+\eta_3\right\}, }
\eqn
\end{widetext}
with a prime denotes the derivative with respect to $r$. By assuming $\Psi = e^{-i \omega t} \Psi(x)$, Eq.~\eqref{master1} could be written as
\bqn
\lb{master2}
\frac{d^2 \Psi(x)}{d x^2}+\left[\omega^2 - V_{\rm eff}(r)\right] \Psi(x) &=& 0.
\eqn
{Here we would like to note that, by introducing  the coordinate $x$, we project the $r\in[r_{MH}, +\infty)$ onto $x\in(-\infty, +\infty)$.  Since the metric  functions, given by Eqs.\eqref{URC_Gr}, \eqref{Gr2} and \eqref{Gr3} behave well on $r\in[r_{MH}, +\infty)$, they will definitely be convergent on  $x\in(-\infty, +\infty)$.}

\begin{table*}
	\caption{Summary of the cases that we will consider for the calculations of QNMs.}  
	\label{table0}
\begin{tabular}{|c c c c c c c c c|}
		\hline
		 & &  & &     & &  & &
		\\[-7pt]
		Case & \quad  \quad & 	Galaxy & \quad  \quad & $G(r)$ & \quad \quad  & $\rho_0~\left(\text{M}_{\odot}/{\rm kpc}^{3} \right)$ & \quad \quad  & $r_0~({\rm kpc})$~~
		\\[1pt]
		\hline
		\hline
		 & &   & &     & &  & &
		\\[-6pt]
		Schwarzschild & &  N/A   & &  $1-\frac{2 M}{r}$ & & $0$ & & N/A
		\\ [2pt]
			\hline
		 & &   & &     & &  & &
		\\[-7pt]
		Case 1 & &  M87  & &  \eqref{URC_Gr}  & & $6.9 \times 10^6$  & &  91.2
		\\
	      & &   Milky Way    & &  \eqref{URC_Gr}  & & $5.2 \times 10^7$  & &  7.8
		\\ [2pt]
		\hline
				 & &   & &     & &  & &
		\\[-7pt]
		Case 2 & &  M87  & & \eqref{Gr2}  & & $0.008 \times10^{7.5}$  & & 130
		\\
	      & &   Milky Way    & &  \eqref{Gr2}  & & $5.23 \times10^7$  & &  8.1
		\\ [2pt]
    	\hline
				 & &   & &     & &  & &
		\\[-7pt]
		Case 3 & &  Milky Way   & &  \eqref{Gr3}  & & $3.43 \times10^7$  & &  15.7
		\\ [2pt]
		\hline
	\end{tabular}
\end{table*}

Following Sec. II, we list the cases that we will consider for the calculations of QNMs in Table \ref{table0} and provide some basic information for each case. They are referred as Case 1, 2 and 3, respectively. The Schwarzschild case is also shown.

\begin{figure}[htb]
	\includegraphics[width=\columnwidth]{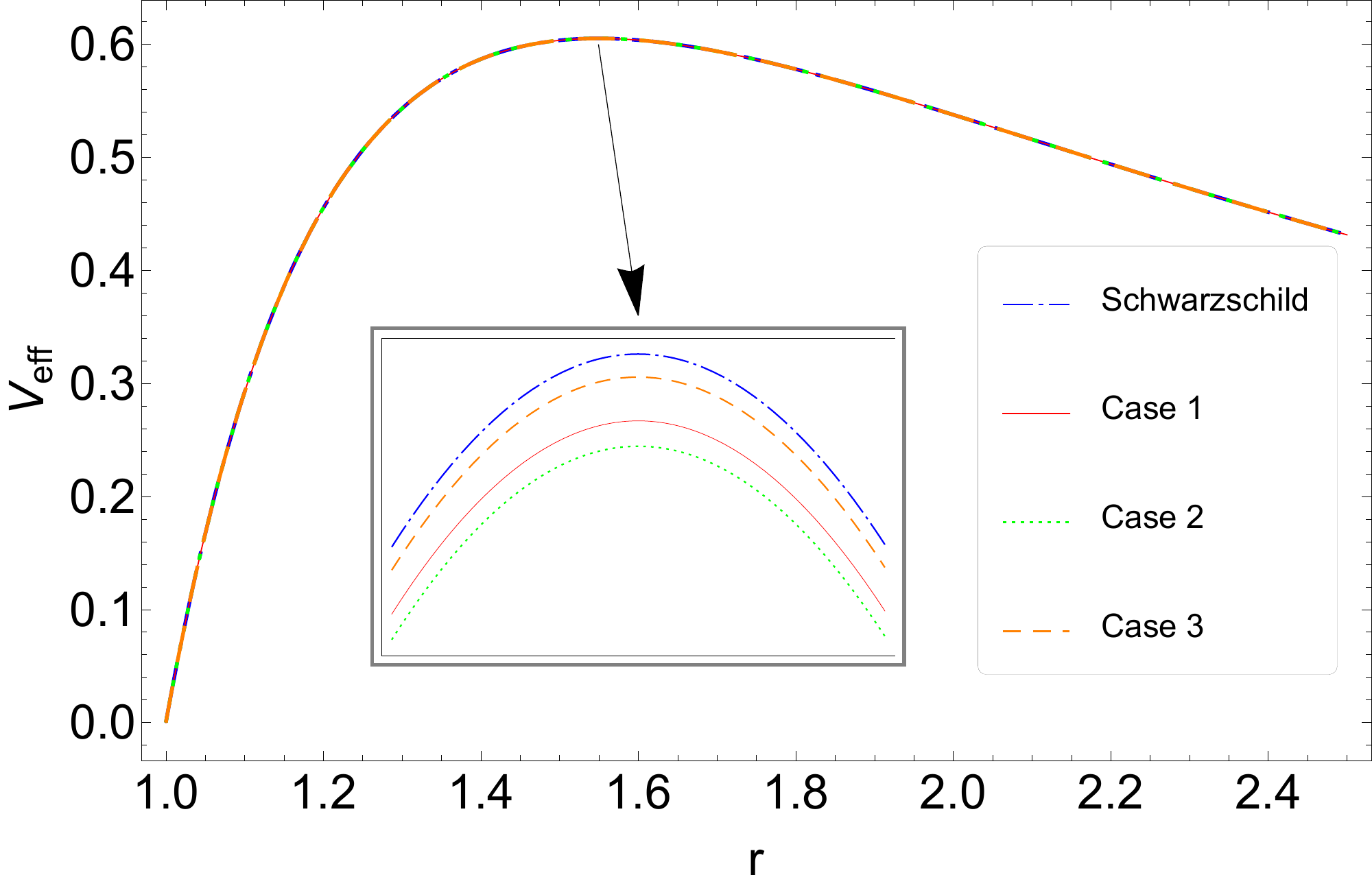}
	\caption{ Behaviors of {$V_{\rm eff}$} for different cases listed in Table \ref{table0}, in which the data for the Milky Way is selected for plotting. In addition, we have chosen $l=2$. {Note that there is an inserted figure showing the amplification of region around the stationary points of these curves.} Also note that here we are using the unit system so that $c=G_N={r_{MH}^{\text{Sch}}}=1$.}
	\label{plot1}
\end{figure}

To find the difference between different cases, $V_{\rm eff}$'s for each case listed in Table \ref{table0} are plotted in Fig. \ref{plot1}. To make them at the same starting line, the data for Milky Way is selected. In addition, as an example, we have set $l=2$ for this plot. Notice that, as notified earlier, here we are using the unit system so that $c=G_N=r_{MH}^{\text{Sch}}=1$. { From Fig. \ref{plot1} it's very clear that the deviation on $V_{\rm eff}$'s between each two cases in Table \ref{table0} is quite small since these curves are almost overlapped. Actually, it is because of this, an amplification of the region around the stationary points of these curves of $V_{\rm eff}$'s is inserted, so that readers can find more details.} That implies we may obtain quite similar QNMs from these cases at the end. As we will see, this is indeed the case.

\section{QNMs of Schwarzshild-like black holes in DM halos}
\renewcommand{\theequation}{4.\arabic{equation}} \setcounter{equation}{0}

With the master equation given by \eqref{master2}, we are ready to solve for the corresponding QNMs, for a specific choice of $G(r)$, including the cases listed in Table \ref{table0} (as well as their extensions by varying $\rho_0$ or $r_0$). Notice that the QNM for cases 2 and 3 have also been considered in \cite{Liu:2021xfb} with larger values of $r_0$ and $\rho_0$. Also notice that, $\omega$ in general is a complex number, often written as $\omega_{l m n}$ \cite{Berti2009}, where $l$ and $m$ are inherited from spherical harmonics while $n$ is the overtone index. However, since we have set $m=0$, it will be left with two indices only, i.e., $l$ and $n$ (For simplicity, before stimulating any confusions, we shall write $\omega_{ln}$ as $\omega$).

\subsection{Calculations of QNMs for the cases in Table \ref{table0}}

Once again, for all the cases mentioned above, we will adopt the unit system so that $c=G_N=r_{MH}^{\text{Sch}}=1$. In this way, the units for mass, time and length are totally fixed.

So far, we have obtained the desired master equation [cf. \eqref{master2}], and we have several background metrics (see Table \ref{table0}). In addition, we also know the two boundary conditions, namely, pure in-going wave at the MH and pure out-going wave at the spatial infinity \cite{Chandra1975}. With everything in hand, and given a set of $\{l, n\}$, we will be able to find the corresponding $\omega_{ln}$.

QNMs in GR with the Schwarzschild case have been studied extensively. In this procedure, several different techniques of calculations were developed. For instance, the  Wentzel-Kramers-Brillouin (WKB) approach \cite{Will1985, Will1987, Konoplya2003, Jerzy2017},  finite difference method (FDM) \cite{XinLi2020}, the continued fraction method \cite{Leaver1985}, shooting method \cite{Chandra1975}, matrix method \cite{Kai2017}, etc.\cite{Kono2011, Gund1994, Bin2004}. Now, we are going to apply some of them to carry out our calculations.
\subsubsection{WKB method}

First of all, we shall try to solve our problem with the sixth-order WKB method. The formula of $\omega$ from the sixth-order WKB method is given by
\bqn
\lb{WKB1}
\omega &=& \sqrt{-i \left[\left(n+\frac{1}{2}\right)+\sum_{k=2}^6 \Lambda_k \right] \sqrt{-2 V_0''}+V_0},~~~~
\eqn
where
\bqn
\lb{WKB2}
V_0 \equiv \left. V_{\rm eff} \right|_{r=r_{\text{max}}}, \quad V_0'' \equiv \left. \frac{d V_{\rm eff}}{d r^2} \right|_{r=r_{\text{max}}},
\eqn
with {$V_{\rm eff} (r=r_{\text{max}})$} gives the maximum of {$V_{\rm eff}$} on $r \in (r_{MH},~\infty)$, where $r_{MH}$ is the radius of the MH. The expressions of $\Lambda_k$'s could be found in \cite{Will1985, Will1987, Konoplya2003}. Note that $n=0, 1, 2, ...$.

\subsubsection{Matrix method}
In fact, we notice that { in using the sixth-order WKB method, our current algorithm will tend to lose its accuracy} when dealing with the $l=2$ case with the polar perturbations. 
Thus, for the $l=2$ case, we shall adopt some new methods. One of them will be referred as the matrix method \cite{Kai2017}. Its basic idea is to discretize \eqref{master2} and put it into a matrix form, so that  we can solve for $\omega$'s by handling an eigenvalue problem. To do so, we first write \eqref{master2} as
\bqn
\lb{master3}
p(r)^2 \Psi''(r)+p(r)p'(r) \Psi'(r)+\left[\omega^2 - V_{\rm eff}(r)\right] \Psi(r) &=& 0, \nb\\
\eqn
where $p(r) \equiv dr/dx$. This equation needs to be solved on $r\in [r_{MH}, +\infty)$ (Notice that, $r_{MH}$ wil be slightly larger than $r_{MH}^{\text{Sch}}$). We shall project this interval to $y \in [0, 1]$ by introducing $y\equiv1-r_{MH}/r$. Thus, \eqref{master3} transforms to
\bqn
\lb{master4}
0 &=& \frac{p(r)^2  (1-y)^4}{r_{MH}^2} \frac{d^2\Psi}{d y^2}  \nb\\
&& + \frac{p(r) [p'(r) r_{MH} - 2 p(r) (1-y)] (1-y)^2}{r_{MH}^2} \frac{d\Psi}{d y} \nb\\
&& +\left[\omega^2 - V_{\rm eff}(r)\right] \Psi.
\eqn

As one can show, the asymptotic solutions of $\Psi(x)$ at the two boundaries, i.e., $r=r_{MH}$ and $r \to \infty$, are given by $\Psi \sim \exp(\pm i \omega x)$. However, to satisfy the purely-in-going and purely-out-going conditions at these two boundaries, we require $\Psi \sim \exp(- i \omega x)$ at the MH and $\Psi \sim \exp(+ i \omega x)$ at the spatial infinity. Such a behavior could be characterized by introducing $\Psi = \mathfrak{F}(y) {\bar \Psi}$. The information of the two physical boundary conditions mentioned above will be encoded in $\mathfrak{F}(y)$ so that we don't have to impose them for ${\bar \Psi}$ $\mathit{ad~hoc} $. In this way, we obtain
\begin{widetext}
\bqn
\lb{master5}
0 &=& \frac{p(r)^2  (1-y)^4}{r_{MH}^2} \frac{d^2 {\bar  \Psi}}{d y^2}  + \frac{p(r) (1-y)^2 \left\{ \left[p'(r) r_{MH} - 2 p(r) (1-y) \right] {\mathfrak{F}} +2 p(r) (1-y)^2 (d{\mathfrak{F}}/dy) \right\}}{r_{MH}^2 {\mathfrak{F} }} \frac{d {\bar  \Psi}}{d y} \nb\\
&& + \left\{\omega^2 - V_{\rm eff}(r) + \frac{\left[p'(r) r_{MH} - 2 p(r) (1-y)\right] (1-y)^2 (d {\mathfrak{F}}/dy) +p(r) (1-y)^4 (d^2 {\mathfrak{F}}/d y^2)}{r_{MH}^2 {\mathfrak{F}} p^{-1} }\right\} {\bar \Psi}.
\eqn
\end{widetext}

 Constructing a suitable ${\mathfrak{F}}(y)$ is not a trivial task in general. For the Schwarzschild case and its simple extensions, we can easily find the corresponding ${\mathfrak{F}}(y)$, as given in \cite{Leaver1985} and  \cite{Kai2017}. However, it can't really solve our problem due to the complicity of $p(r)$. To conquer this problem, we shall explore a kind of approximate technique.

 \begin{figure}[htb]
	\includegraphics[width=\columnwidth]{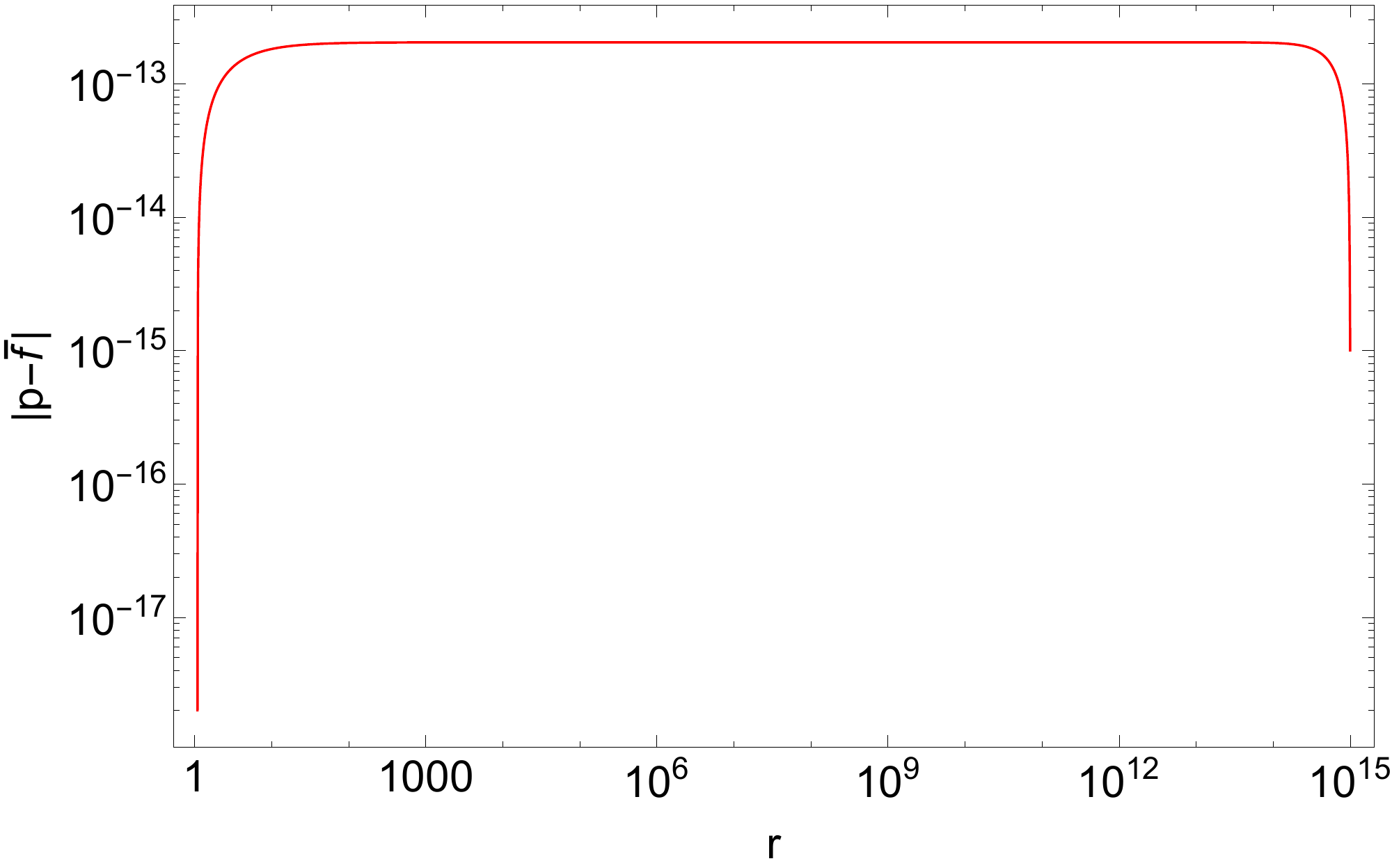}
	\caption{ Amplitude of $(p-{\bar f})$ on $r\in (r_{MH}, 10^{15})$ for  Case 3 by setting $l=2$, $r_0=15.7 \text{kpc}$ and $\rho_0 = 3\times 10^{12} \text{M}_{\odot}/{\rm kpc}^{3}$. Note that here we are using the unit system so that $c=G_N={r_{MH}^{\text{Sch}}}=1$.}
	\label{plot5}
\end{figure}

 To illustrate the detail, we use Case 3 in Table \ref{table0} as our example (so that the data for the Sgr $\text{A}^\ast$ BH will be used). In that case, we choose $l=2$, $r_0=15.7 \text{kpc}$ and $\rho_0 = 3\times 10^{12} \text{M}_{\odot}/{\rm kpc}^{3}$ (Here we select a sufficiently large $\rho_0$ so that the following analysis will be valid even when our problem deviates significantly from the Schwarzschild case). After that, we notice that the  corresponding $p(r)$ could be well approximated by
 \bqn
 \lb{barf}
 {\bar f} (r) \equiv {\bar \alpha} (1-r_{MH}/r),
 \eqn
 near the MH as well  as the spatial infinity, where ${\bar \alpha}$ is a factor to be determined. For the Schwarzschild case, this is of course the truth. Given a DM halo, the validity of such an approximation could be justified by Fig. \ref{plot5}, in which  the  amplitude of  $(p-{\bar f})$ is shown on $r\in (r_{MH}, 10^{15})$. In this interval, we observe that the amplitude of $(p-{\bar f})$ stays extremely small, which makes our approximation legal. In fact, by definition, we have $p(r=r_{MH})={\bar f}(r=r_{MH})=0$, so it's not surprising that $p(r)$ could be well approximated by ${\bar f}(r)$ near the MH.  On the other hand, according to \eqref{rho3}, the density of the DM behaves like $\rho \sim {\cal{O}} (r^{-1})$ near the spatial infinity, which means the influence of DM will rapidly fade away. Thus, it's reasonable to treat an $r$ as large as $r=10^{15}$ to be the cutoff point for the influence of DM. In other words, ${\bar f}(r)$ could sufficiently approximate $p(r)$ on $r\in(10^{15}, +\infty)$ due to the lack of DM on this region. In this way, we have proven that we are safe to use ${\bar f}(r)$ to approximate $p(r)$ near the MH as well as the spatial infinity. Thus, we have $x \approx [r+r_{MH} \ln (r- r_{MH})]/{\bar \alpha}$. As a result, by mimicking \cite{Kai2017}, the ${\mathfrak{F}}(y)$ could be constructed  as
 \bqn
\lb{goF}
{\mathfrak{F}}(y) &=& \frac{e^{(i \omega r_{MH})/[{\bar  \alpha}(1-y)]} }   { (1-y)^{i \omega r_{MH}/{\bar \alpha}} ( y r_{MH})^{i \omega r_{MH}/{\bar \alpha}}}.
\eqn
As one can find, ${\mathfrak{F}}(y)$ in \eqref{goF} is indeed the ${\mathfrak{F}}(y)$ that guarantees the two physical boundary conditions of $\Psi$ mentioned earlier. It needs to be emphasized here that the approximation of $p(r)$ will only be used to construct  ${\mathfrak{F}}(y)$. The $p(r)$ in \eqref{master5} will be evaluated with its exact expression.

After that, by following \cite{Kai2017}, we introduce ${\hat \Psi} = y(1-y) {\bar \Psi}$, and obtain
\begin{widetext}
\bqn
\lb{master6}
0 &=& \frac{p(r)^2  (1-y)^3}{r_{MH}^2 y} \frac{d^2 {\hat  \Psi}}{d y^2}  + \frac{p(r) (1-y) \left\{ \left[p'(r) r_{MH} y - 2 p(r) (1-y)^2 \right] {\mathfrak{F}} +2 p(r) (1-y)^2 y (d{\mathfrak{F}}/dy) \right\}}{r_{MH}^2 y^2 {\mathfrak{F} }} \frac{d {\hat  \Psi}}{d y} \nb\\
&& + \left\{\omega^2  + \frac{(1-y)y\left[ \left(p'(r) r_{MH} y - 2 p(r) (1-y)^2 \right)(d {\mathfrak{F}}/dy) +p(r) (1-y)^2 y (d^2 {\mathfrak{F}}/d y^2)\right]   }{r_{MH}^2 {\mathfrak{F}} p^{-1} y^2/(1-y) }\right\} y(1-y) {\hat \Psi} \nb\\
&& + \left\{- V_{\rm eff}(r) + \frac{ \left[2 p(r) (1-y)^3-p'(r) r_{MH} (1- 2y) y\right] {\mathfrak{F}}  }{r_{MH}^2 {\mathfrak{F}} p^{-1} y^2/(1-y) }\right\} y(1-y) {\hat \Psi}.
\eqn
\end{widetext}
In practice, we shall carry out our calculations with \eqref{master6}. To apply the matrix  method to \eqref{master6}, we need to discretize the variable $y$ by replacing it with $y_i = (i-1)/(N-1)$, where $i \in [1, N]\cap \mathbb{Z}$, with $N$ a positive integer (It will be chosen properly to meet our tolerance of accuracy. For instance, in this paper { we set $N=22$).} In this way, $y_i$'s are distributing evenly on the interval $[0, 1]$. Similarly, we discretize ${\hat \Psi}$ by replacing it with a set of ${\hat \Psi}_i$, where $i \in [1, N]\cap \mathbb{Z}$.

At the same time, we introduce a set of $(N-1)$-dimension vectors ${\vec{\mathfrak{V} }}_i = ({\hat \Psi}_1-{\hat \Psi}_i, {\hat \Psi}_2-{\hat \Psi}_i, ..., {\hat \Psi}_{i-1}-{\hat \Psi}_i, {\hat \Psi}_{i+1}-{\hat \Psi}_i, ..., {\hat \Psi}_{N-1}-{\hat \Psi}_{i}, {\hat \Psi}_N-{\hat \Psi}_i)$ as well as a set of $(N-1)\times (N-1)$ matrices ${\mathbb{M}}_i$ with $({\mathbb{M}}_i)_{jk} = (y_j - y_i)^k/(j!)$, where $i, j\in [1, N]\cap \mathbb{Z}$ with $i\ne j$ and $k \in [1, N-1]\cap \mathbb{Z}$. By using Taylor's expansion law on $\hat \Psi$ around $y_i$'s, we  have ${\vec{\mathfrak{D}}}_i = {\mathbb{M}}_i^{-1} \cdot {\vec{\mathfrak{V} }}_i$ (Note that here we are not using Einstein’s summation convention \cite{Gronbook}),  where ${\vec{\mathfrak{D}}}_i$ represents a set of $(N-1)$-dimension vectors given by ${\vec{\mathfrak{D} }}_i = ({\hat \Psi}^{(1)}(y_i), {\hat \Psi}^{(2)}(y_i), {\hat \Psi}^{(3)}(y_i), ..., {\hat \Psi}^{(N-2)}(y_i), {\hat \Psi}^{(N-1)}(y_i))$, with  the $(n)$ in the superscripts meaning the $n$th derivative {with respect to their  arguments.} Thus, we can solve  for ${\hat \Psi}^{(1)}(y_i)$ as well as ${\hat \Psi}^{(2)}(y_i)$ and substitute  them into \eqref{master6} for $d {\hat \Psi}/dy$ as well as $d^2 {\hat \Psi}/dy^2$, respectively. In this way, we accomplish  the discretization of \eqref{master6}. As a result, we obtain $N$ linear equations of ${\hat \Psi}_i$'s. They could be  expressed in a matrix form as ${\vec 0} = {\mathbb{K}} \cdot {\vec{\mathfrak{W}}} $, where ${\vec{\mathfrak{W}}}$ is an $N$-dimension vector given by ${\vec{\mathfrak{W}}} = ({\hat \Psi}_1, {\hat \Psi}_2, ..., {\hat \Psi}_N)$. Here, the ${\mathbb{K}}$ is an $N\times N$ matrix and its components are read off from the $N$ linear  equations mentioned  above.

On top of that, we further construct
${\vec 0} ={ \bar {\mathbb{K}}} \cdot {\vec{\mathfrak{W}}}$,
 where ${ \bar {\mathbb{K}}}_{ij} = {\mathbb{K}}_{ij} (1-\delta_{1, i})(1-\delta_{N, i}) + \delta_{1, i} \delta_{i, j} + \delta_{N, i} \delta_{i, j}$  (Note that here we are not using Einstein’s summation convention either). Clearly, this is nothing but an eigenvalue problem. Thus, we can solve for $\omega$'s through
\bqn
\lb{detK}
\det{({ \bar {\mathbb{K}}})} &=& 0.
\eqn
The above procedures as well as analysis could be repeated for different cases as well as different $\rho_0$'s and $r_0$'s.


\begin{table*}
	\caption{The QNMs $\omega$ for BHs with DM halo by adopting the background and factors given in Case 1 of Table \ref{table0}. Note that here $\omega$'s from both polar and axial perturbations are compared with their Schwarzschild counterparts (Recall the isospectrality in the Schwarzschild case).}  
	\label{table1}
\begin{tabular}{|c c c c | c c c c | c c c c | c|}
		\hline
		 & \quad   & & \quad   & Axial perturbations: & \quad \quad &   & \quad \quad  &  Polar perturbations: & \quad \quad &  & \quad \quad  &  Axial$\setminus$Polar:
		\\
		\hline
		$l$ & \quad   & $n$ & \quad   &  M87 & \quad \quad & Milky Way & \quad \quad  &  M87 & \quad \quad & Milky Way & \quad \quad  &  Schwarzschild case~
		\\
		\hline
		\hline
		2  & & 0 & &  $ 0.74728 	-0.17791 i $   & & $ 0.74734 	-0.17792 i $ & &  $ 0.74728 	-0.17791 i $   & &  $ 0.74734 	-0.17792 i $   & &  $ 0.74734 	-0.17792 i $
		\\
		  & & 1 & &  $ 0.69330 	-0.54777 i $   & & $ 0.69335 	-0.54782 i $ & & { $ 0.69335 	-0.54779 i $ }  & & { $ 0.69340 	-0.54784 i $ }  & &  $ 0.69342 	-0.54783 i $
		\\
		\hline
		3  & & 0 & &  $1.19879 	-0.18539i$   & & $ 1.19888 	-0.18540i $ & &  $1.19879 	-0.18539 i$   & &  $1.19888 	-0.18540 i$   & &  $1.19889 	-0.18541i$
		\\
		   & &   1   & &  $1.16519 	-0.56252i$  & & $ 1.16528 	-0.56258i $   & &  $1.16519 	-0.56253 i$   & &  $1.16528 	-0.56258 i$   & &  $1.16528 	-0.56258i$
		\\
       \hline
		4  & &   0   & &  $1.61823 	-0.18831i$  & & $ 1.61835 	-0.18833i $   & &  $1.61823 	-0.18831 i$   & &  $1.61835 	-0.18833 i$   & &  $1.61836 	-0.18833i$
		\\
		   & &   1   & &  $1.59313 	-0.56861i$  & & $ 1.59326 	-0.56866i $  & &  $1.59313 	-0.56861 i$   & &  $1.59326 	-0.56866 i$   & &  $1.59326 	-0.56867i$
		\\
   \hline
		5  & &   0   & &  $ 2.02443 	-0.18972 i $  & &  $ 2.02458 	-0.18974 i $   & &  $ 2.02443 	-0.18972 i $   & &  $ 2.02458 	-0.18974 i $   & &  $ 2.02459 	-0.18974 i $
		\\
		   & &   1   & &  $ 2.00428 	-0.57157 i $  & & $ 2.00443 	-0.57163 i $  & &  { $ 2.00443 	-0.57163 i $ }   & &  $ 2.00443 	-0.57163 i $   & &  $ 2.00444 	-0.57163 i $
	\\
       \hline
	\end{tabular}
\end{table*}

\begin{table*}
	\caption{The QNMs $\omega$  for BHs with DM halos by adopting the background and factors given in Case 2 of Table \ref{table0}. Note that here $\omega$'s from both polar and axial perturbations are compared with their Schwarzschild counterparts (Recall the isospectrality in the Schwarzschild case).}  
	\label{table2}
\begin{tabular}{|c c c c | c c c c | c c c c | c|}
		\hline
		 & \quad   & & \quad   & Axial perturbations: & \quad \quad &   & \quad \quad  &  Polar perturbations: & \quad \quad &  & \quad \quad  &  Axial$\setminus$Polar:
		\\
		\hline
		$l$ & \quad   & $n$ & \quad   &  M87 & \quad \quad & Milky Way & \quad \quad  &  M87 & \quad \quad & Milky Way & \quad \quad  &  Schwarzschild case~
		\\
		\hline
		\hline
		2  & & 0 & &  $ 0.74734 	-0.17792 i $   & & $ 0.74734 	-0.17792 i $ & &  $ 0.74734 	-0.17792 i $   & &  $ 0.74734 	-0.17792 i $   & &  $ 0.74734 	-0.17792 i $
		\\
		  & & 1 & &  $ 0.69335 	-0.54782 i $   & & $ 0.69335 	-0.54782 i $ & &  $ 0.69339 	-0.54784 i $   & &  { $ 0.69339 	-0.54784 i $ }  & &  $ 0.69342 	-0.54783 i $
		\\
		\hline
		3  & & 0 & &  $1.19888 	-0.18540i$   & & $ 1.19888 	-0.18540i $  & &  $1.19888 	-0.18540 i$   & &  $1.19888 	-0.18540 i$    & & $1.19889 	-0.18541i$
		\\
		   & &   1   & & $1.16528 	-0.56258 i$   & & $ 1.16528 	-0.56258i $  & &   $1.16528 	-0.56258 i$  & &  $1.16528 	-0.56258 i$    & & $1.16528 	-0.56258i$
		\\
       \hline
		4  & &   0   & & $1.61834 	-0.18833 i$   & & $ 1.61835 	-0.18833i $  & &  $1.61834 	-0.18833 i$   & &  $1.61835 	-0.18833 i$    & & $1.61836 	-0.18833i$
		\\
		   & &   1   & & $1.59325 	-0.56866 i$   & & $ 1.59325 	-0.56866i $  & &  $1.59325 	-0.56866 i$   & &  $1.59325 	-0.56866 i$    & & $1.59326 	-0.56867i$
		\\
       \hline
		5  & &   0   & &  $ 2.02458 	-0.18974 i $  & &  $ 2.02458 	-0.18974 i $   & &  $ 2.02458 	-0.18974 i $   & &  $ 2.02458 	-0.18974 i $   & &  $ 2.02459 	-0.18974 i $
		\\
		   & &   1   & &  $ 2.00443 	-0.57163 i $  & & $ 2.00443 	-0.57163 i $  & &  $ 2.00443 	-0.57163 i $   & &  $ 2.00443 	-0.57163 i $   & &  $ 2.00444 	-0.57163 i $
	\\
       \hline
	\end{tabular}
\end{table*}

\begin{table*}
\centering
	\caption{The QNMs $\omega$ for BHs with DM halos by adopting the background and factors given in Case 3 of Table \ref{table0}. Note that here $\omega$'s from both polar and axial perturbations are compared with their Schwarzschild counterparts (Recall the isospectrality in the Schwarzschild case).}  
	\label{table3}
\begin{tabular}{|c c c c | c c c c | c c c c | c|}
		\hline
		 & \quad   & & \quad   & Axial perturbations: & \quad \quad &   & \quad \quad  &  Polar perturbations: & \quad \quad &  & \quad \quad  &  Axial$\setminus$Polar:
		\\
		\hline
		$l$ & \quad   & $n$ & \quad   &  M87 & \quad \quad & Milky Way & \quad \quad  &  M87 & \quad \quad & Milky Way & \quad \quad  &  Schwarzschild case~
		\\
		\hline
		\hline
		2  & & 0 & & N/A   & & $  0.74734 	-0.17793 i $ & & N/A   & &  $  0.74734 	-0.17792 i $   & &  $ 0.74734 	-0.17792 i $
		\\
		  & & 1 & & N/A   & & $ 0.69335 	-0.54782 i $ & &  N/A   & &  { $ 0.69340 	-0.54785 i $ }  & &  $ 0.69342 	-0.54783 i $
		\\
       \hline
		3  & & 0 & &   N/A  & & $ 1.19888 	-0.18540i $  & &  N/A  & &   $1.19888 	-0.18541 i$   & &  $1.19889 	-0.18541i$
		\\
		   & &   1   & &  N/A  & & $ 1.16528 	-0.56258i $  & &  N/A  & &   $1.16528 	-0.56258 i$   & &  $1.16528 	-0.56258i$
		\\
       \hline
		4  & &   0   & &  N/A  & & $ 1.61835 	-0.18833i $  & & N/A   & &  $1.61835 	-0.18833 i$    & &  $1.61836 	-0.18833i$
		\\
		   & &   1   & &  N/A  & & $ 1.59326 	-0.56867i $  & &  N/A  & &   $1.59326 	-0.56867 i$   & &  $1.59326 	-0.56867i$
		\\
       \hline
		5  & &   0   & &  N/A  & &  $ 2.02459 	-0.18974 i $   & &  N/A   & &  $ 2.02459 	-0.18974 i $   & &  $ 2.02459 	-0.18974 i $
		\\
		   & &   1   & &  N/A  & & $ 2.00444 	-0.57163 i $  & &  N/A   & &  $ 2.00444 	-0.57163 i $   & &  $ 2.00444 	-0.57163 i $
	\\
       \hline
	\end{tabular}
\end{table*}

Knowing the above methods, we are ready to run our calculations of $\omega$'s. In this subsection, we shall adopt the choices of parameters given in Table \ref{table0} to carry out the calculations. The results of $\omega$'s (for $l=2,\; 3,\;4,\; 5$ and $n=0,\;1$) for this part are exhibited in Tables \ref{table1} - \ref{table3} for Case 1, 2 and 3, respectively. Note that in these tables, the results for BHs with DM halo are compared with their counterparts from the Schwarzschild case and the axial perturbations \cite{Chao2021}. Also note that the results for both axial and polar perturbations in the Schwarzschild case  are identical due to the isospectrality \cite{Berti2009, Chand83} in this case.

By looking at Tables \ref{table1} - \ref{table3}, we immediately notice that the deviations between the Schwarzschild and non-Schwarzschild cases occur at the 4th digit or after that. For all the listed  $\omega$'s in Tables \ref{table1} - \ref{table3}, these deviations are very small, just like we anticipated earlier. Considering the fact that our calculations contain numerical errors, these deviations are quite negligible.
On the other hand, by comparing the results from the axial and polar perturbations, we can confirm their isospectrality up to about the 5th digit.

\subsection{QNMs for the $l=2$ case with different values of $\rho_0$ and $r_0$}

In the above subsection, we have considered the impacts of DM halos on QNMs by adopting the halo parameters $\rho_0$ and $r_0$ given in Table.~\ref{table0}. It is worth mentioning here that these parameters are in general derived by fitting the corresponding density profiles with the observational data of the rotation curves in different galaxies, see \cite{dark_matter} for examples. Thus, these profiles roughly reflect DM distributions for the whole galaxy. They tend to be accurate in describing the regions that far away from the central BH. In contrast, to the contexts of matter environment around the central BH, the halo parameters $\rho_0$ and $r_0$ are basically free.

In addition, the values of $\rho_0$ and $r_0$ also change from galaxy to galaxy. In the Milky Way, $r_0 \sim 10 \;{\rm kpc}$ and $\rho_0 \sim 10^{7} M_{\odot}/{\rm kpc}^{3}$, as presented in Table.~\ref{table0}. Depending on specific galaxies, it is shown that in the catalog of dark matter halo models for galaxies in the Spitzer Photometry and Accurate Rotation Curves (SPARC) database, $r_0$ can be as large as $ \sim 700\; {\rm kpc}$ and $\rho_0$ can be as large as $10^{10} M_{\odot}/{\rm kpc}^{3}$ for NFW profile \cite{Li:2020iib}. In the SPARC, 175 galaxies are included. Thus, it is natural to expect that the values of $\rho_0$ or $r_0$ in some galaxies in the Universe could be even larger than those given in SPARC.

 \begin{figure}[htb]
	\includegraphics[width=\columnwidth]{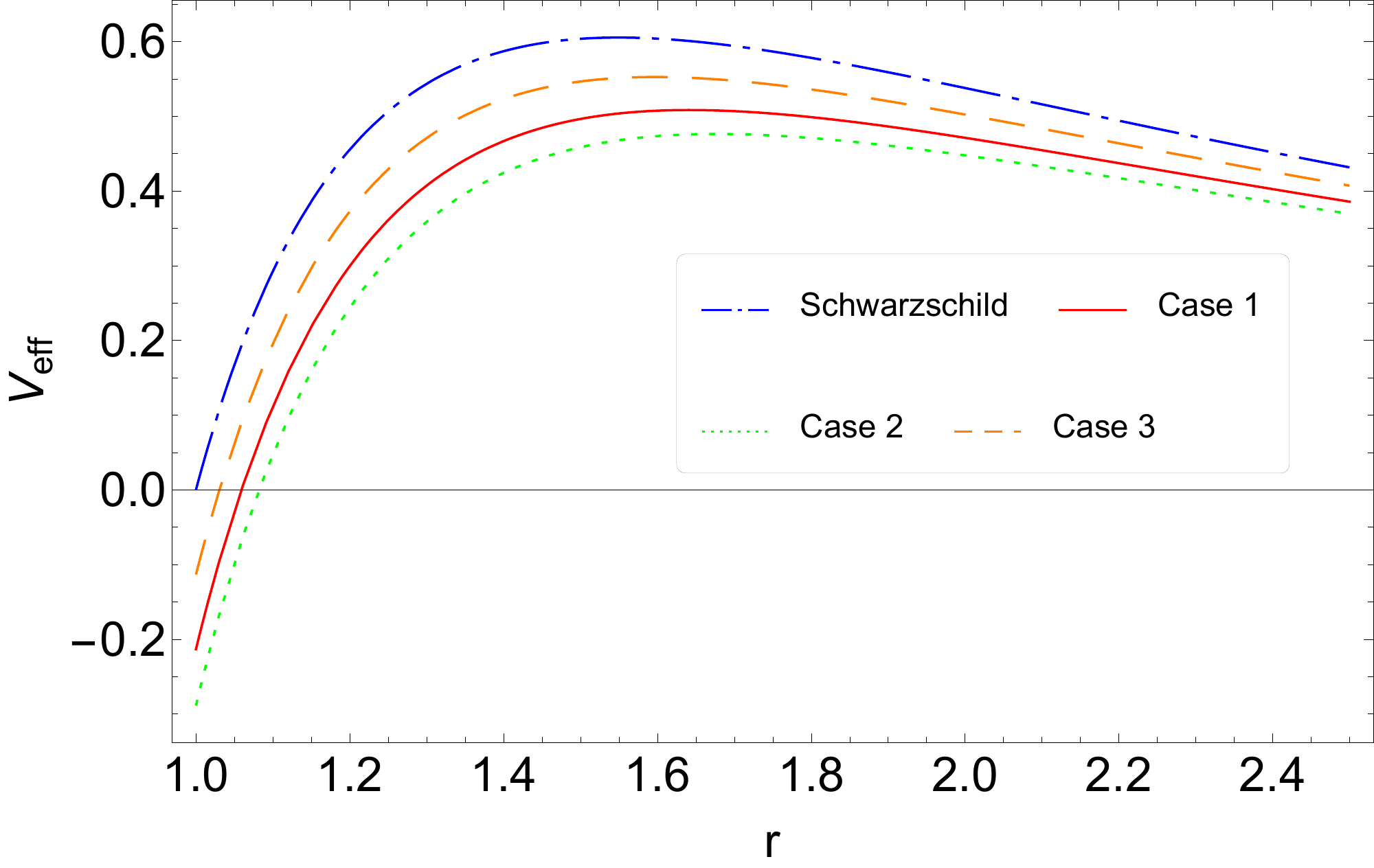}
	\caption{ Behaviors of {$V_{\rm eff}$} for different cases listed in Table \ref{table0} by setting $l=2$. Here we have chosen $\rho_0=10^{12} \text{M}_{\odot}/{\rm kpc}^{3}$ and the values for the other parameters are from the data for the Milky Way. Note that here we are using the unit system so that $c=G_N={r_{MH}^{\text{Sch}}}=1$.}
	\label{plot4}
\end{figure}

For these reasons, it's worth investigating how the frequencies of QNMs shift with different values of $\rho_0$ and $r_0$.
Thus, taking this opportunity, we also test the influence of $\rho_0$ and $r_0$ on QNMs. At the same time, since the $l=2$ mode is in general the dominate one \cite{Mag18}, in this part we shall focus ourselves on this mode only.

\begin{figure}[htb]
	\includegraphics[width=\columnwidth]{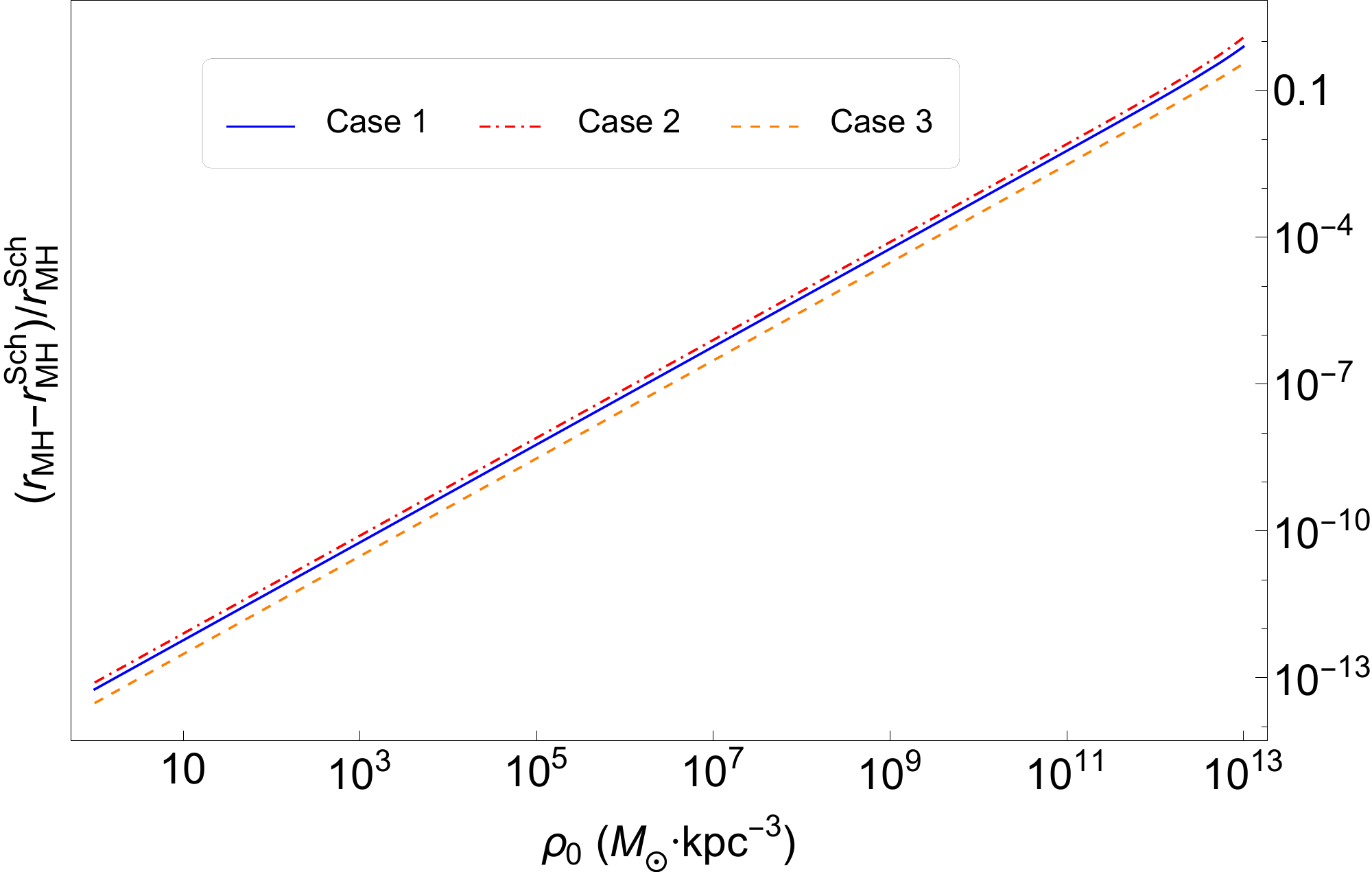}
	\includegraphics[width=\columnwidth]{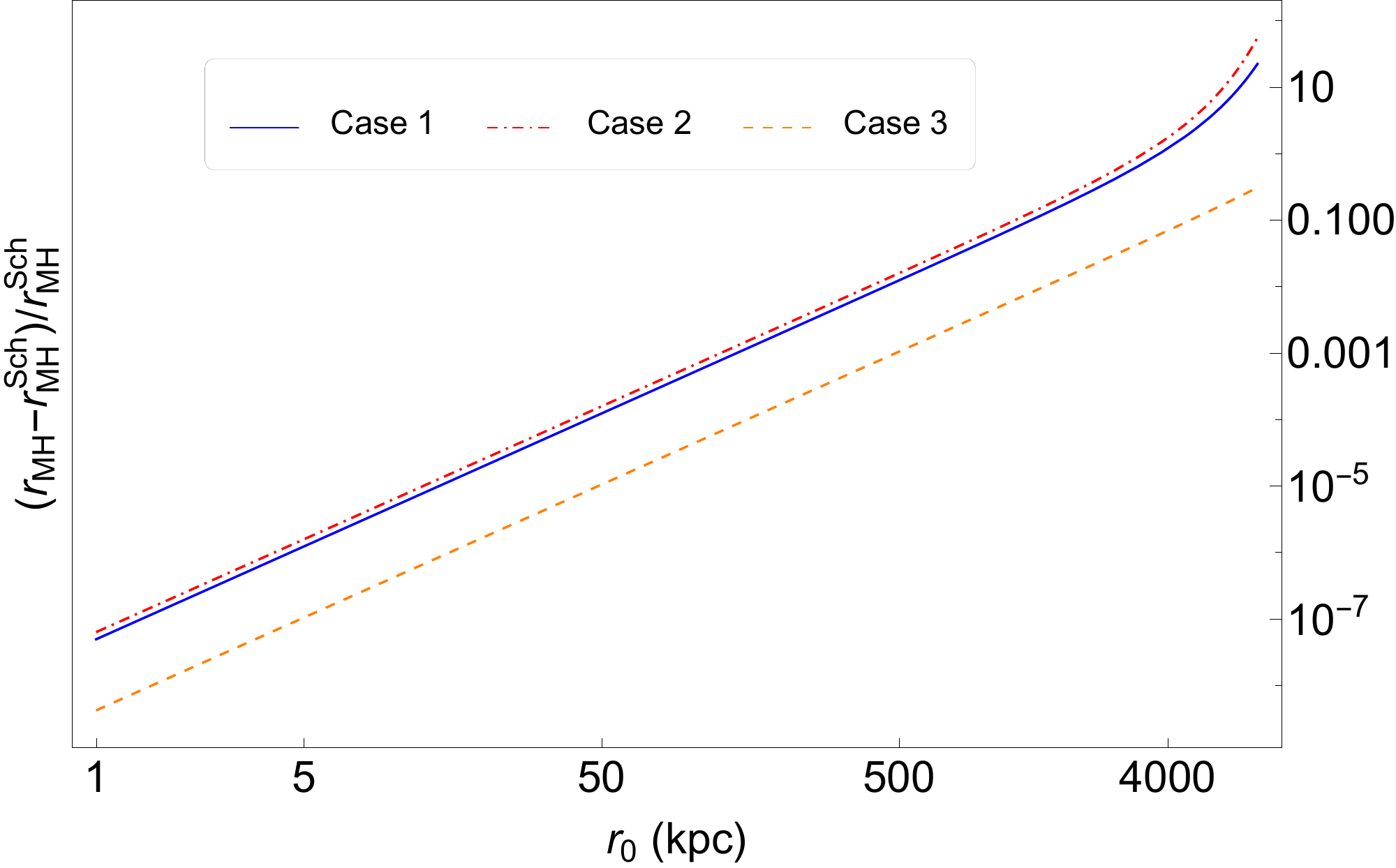}
	\caption{The relative difference between  $r_{MH}$ and $r_{MH}^{\text{Sch}}$ as functions of $\rho_0$ and $r_0$ for  the three cases  listed in Table \ref{table0}. In the  upper panel, the $r_0$'s for these three cases are picked up from the Milky Way data in Table \ref{table0}.   In the  lower panel, the $\rho_0$'s for these three cases are picked up from the  Milky Way data in Table \ref{table0}}
	\label{plot7}
\end{figure}

Since the final results are directly related to the effective potential, we first take a look to that. In Fig. \ref{plot4} we show the behaviors of $V_{\rm eff}$ for the three cases listed in Table \ref{table0} by setting $l=2$ and $\rho_0=10^{12} \text{M}_{\odot}/{\rm kpc}^{3}$. Just like in Fig. \ref{plot1}, all the other parameters are chosen from the data for the Milky Way (cf. Table \ref{table0}). By looking at Fig. \ref{plot4} and comparing it with Fig. \ref{plot1}, we notice that, for the curves of $V_{\rm eff}$'s in all these three cases, the regions around the stationary points will be more like plateaux (instead of peaks) when we have a larger $\rho_0$. Another phenomenon is that the position of a $V_{\rm eff}$'s root moves a little bit to the right when we have a larger $\rho_0$. {Essentially, that's because the radius of the MH is getting bigger. To show this more clear, we plot out the relative differences between $r_{MH}$ and $r_{MH}^{\text{Sch}}$ as functions of $\rho_0 \setminus r_0$ for the three cases in Fig \ref{plot7} (The Milky Way data is used in there). From  there we find it clearly that, given
sufficiently large $\rho_0$'s or $r_0$'s (in comparing to the  ones listed in Table \ref{table0}), the resultant $r_{MH}$ can significantly deviate from $r_{MH}^{\text{Sch}}$.} That implies that we may find non-trivial discrepancies between the Schwarzschild and non-Schwarzschild cases based on QNMs with a large enough $\rho_0$ (the same for $r_0$), as we have seen in \cite{Chao2021}. In addition, we also anticipate that all the three cases listed in Table \ref{table0} will share similar patterns when we adjust $\rho_0$ or $r_0$. Thus, we shall simply take Case 3 as our example.

As mentioned in the last subsection, the $l=2$ case needs to be handled more carefully. Therefore, in this subsection the FDM \cite{XinLi2020, Habermanb} will be used for the calculation. Different from the WKB approach and the matrix method, using the FDM, we shall solve \eqref{master1} for $\Psi$ (which carries the information of $\omega_{ln}$'s) in the time domain. An advantage of using this method here is that in this way we can observe the discrepancies mentioned above clearly.

To apply the FDM, we first introduce two new variables $\mu \equiv t-x$ and  $\nu \equiv t+x$ [so that  $t=(\nu+\mu)/2$ and $x=(\nu-\mu)/2$]\footnote{ {One of the biggest differences between the FDM and the other two methods mentioned above is that in using the FDM, we need the exact form of $x(r)$, in addition to its derivative with respect to $r$. Therefore, according to the definition \eqref{rast}, we have to assign $x$ an integral constant. In fact, such a constant could be chosen arbitrarily and it's independent of our results for $\omega$'s. Thus, we made a simple choice by letting $x (r=2)=2$. }}. Therefore, on a $(N+1) \times (N+1)$ lattice (where $N$ is a positive integer that will be chosen properly according to our usage), we perform the calculation of $\Psi(\mu, \nu)$ by using {the recursion formula}
\begin{widetext}
\bqn
\lb{FDM1}
\Psi(\mu+\delta h, \nu+\delta h) &\cong & \Psi(\mu, \nu+\delta h) + \Psi(\mu+\delta h, \nu) - \Psi(\mu, \nu )-\delta h^2 V_{\rm eff}\left(r \right) \frac{\Psi(\mu, \nu+\delta h ) + \Psi(\mu+\delta h, \nu)}{8},
\eqn
\end{widetext}
where $\delta h$ is the step size. The boundary conditions are given by $ \Psi(\mu, \nu=0 ) = 0\;(\mu \ne 0)$ and $ \Psi(\mu=0, \nu ) = \exp[-(\nu-1)^2/2]$. Thus, after $N^2$ iterations, we find all the $\Psi(n \delta h, n \delta h)$'s for $n \in [0, N]\cap \mathbb{Z}$. From that we can calculate $\Psi(t, x=0 )$ by using the relation $\Psi(t, x=0 )=\Psi(\mu, \mu)$.

\begin{figure}[htb]
	\includegraphics[width=\columnwidth]{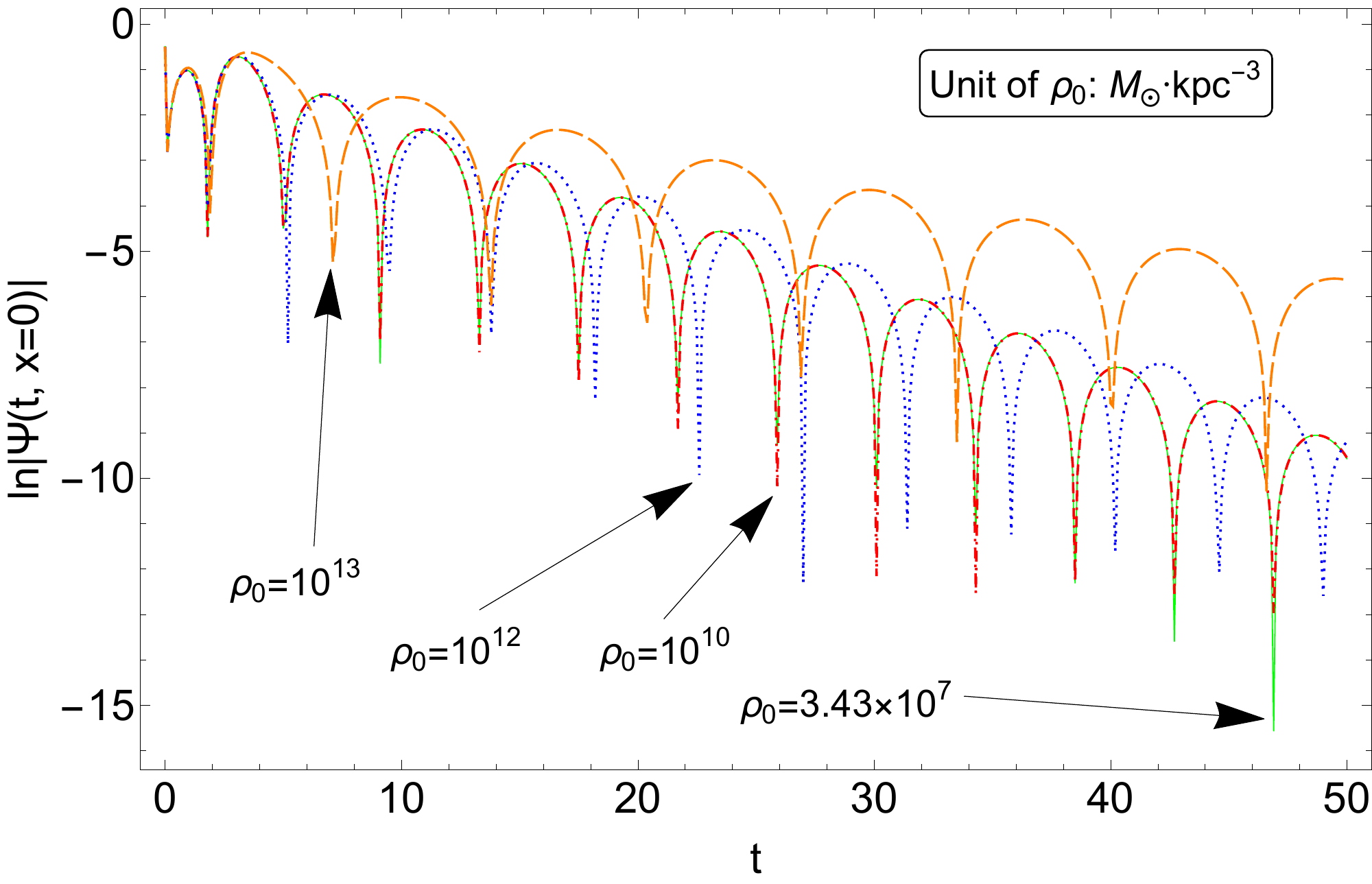}
	\includegraphics[width=\columnwidth]{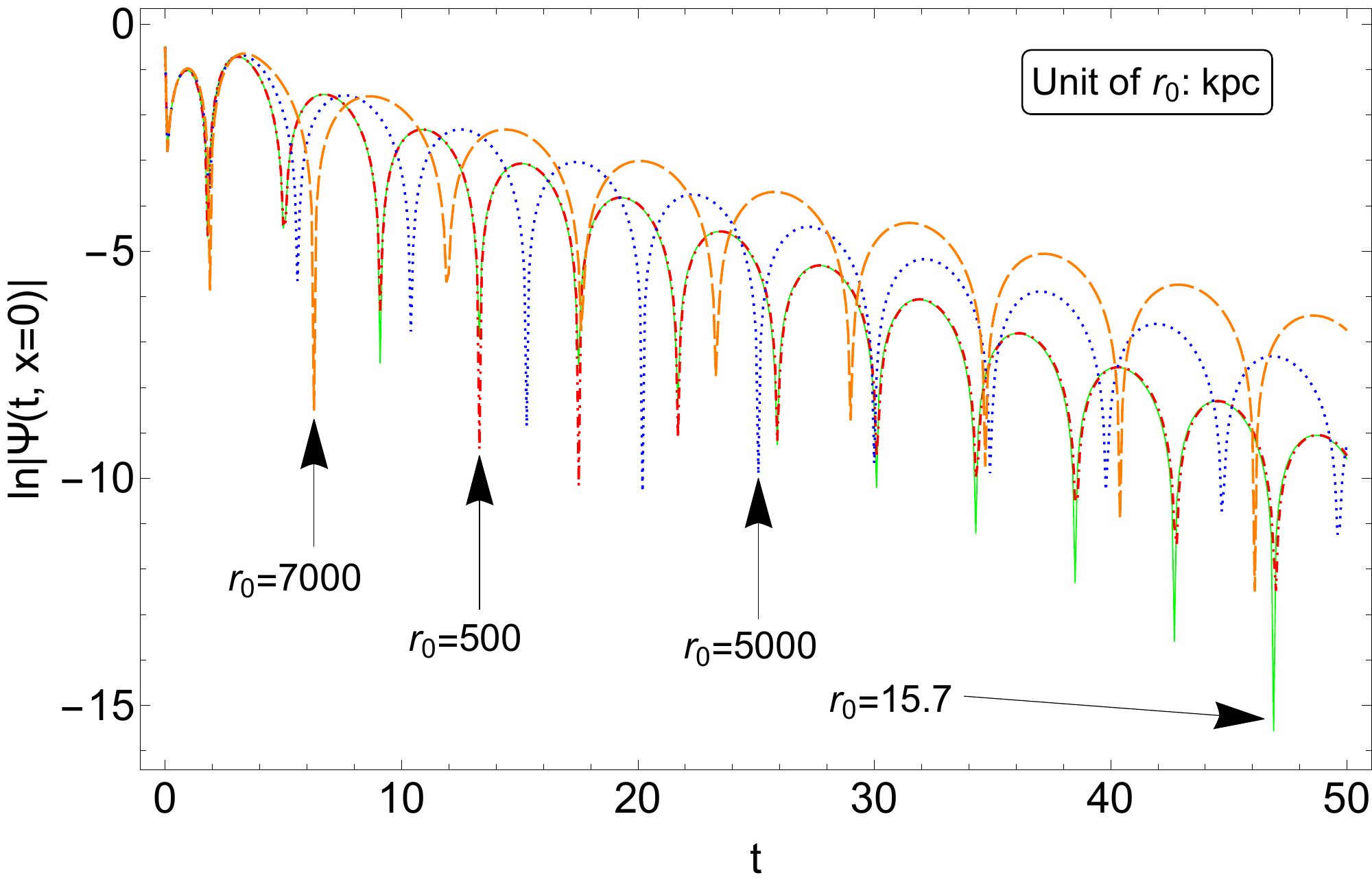}
	\caption{ The temporal evolution of $\Psi(t, x=0)$ [cf. \eqref{master1}] for Case 3 in Table. \ref{table0} by setting $l=2$. For the upper panel, we fix $r_0=15.7 \text{kpc}$. For the lower panel, we fix $\rho_0=3.43\times10^{7} \text{M}_{\odot}/{\rm kpc}^{3}$. In the upper panel, the green solid line, red dash-dotted line, blue dotted line and orange dashed line represent the results for $\rho_0=3.34\times10^7$, $\rho_0=10^{10}$,  $\rho_0=10^{12}$ and  $\rho_0=10^{13}$ (in $\text{M}_{\odot}/{\rm kpc}^{3}$), respectively. In the lower panel, the green solid line, red dash-dotted line, blue dotted line and orange dashed line represent the results for $r_0=15.7$, $r_0=500$,  $r_0=5000$ and  $r_0=7000$ (in ${\rm kpc}$),  respectively. Note that here we are using the unit system so that $c=G_N={r_{MH}^{\text{Sch}}}=1$.}
	\label{plot2}
\end{figure}

In Fig. \ref{plot2} we plot out $\ln |\Psi(t, x=0)|$,
where $t\in [0, 50]$, for Case 3 by setting  $l=2$. For the upper panel, we fix $r_0=15.7 \text{kpc}$ and vary $\rho_0$, while for the lower panel, we fix $\rho_0=3.34\times10^7 \text{M}_{\odot}/{\rm kpc}^{3}$ and vary $r_0$.  During the calculations, we choose $N=500$ and $\delta h=0.1$. Notice that, in there the shapes of these curves reflect the comprehensive effects of all the existing $\omega_{2n}$'s (Of course, in principle, $\omega_{20}$ is the dominate one). Roughly speaking, the slopes of these curves' fitting lines represent  $\omega_{2n}$'s imaginary parts (the damping time) while their periods represent $\omega_{2n}$'s real parts (the periods of vibration) \footnote{ In fact, we are able to extract the exact value of each $\omega$ from the FDM by using, e.g., the Prony method \cite{Berti2007}. However, this work won't be straightforward and more importantly, to our knowledge, it's not  easy to control the  accuracy of the Prony method. Thus, we are not going to demonstrate the detail of Prony method here. }.

Knowing this and looking at the upper panel of Fig. \ref{plot2}, we find that when the $\rho_0$ is apart from $3.34\times10^7 \text{M}_{\odot}/{\rm kpc}^{3}$ and getting bigger, the changing rate of the resultant $\omega_{2n}$'s is very small at the beginning and will increase significantly when $\rho_0$ is big enough. Similarly, by looking at the lower panel of Fig. \ref{plot2}, we find that when $r_0$ is apart from $15.7{\rm kpc}$ and getting bigger, the changing rate of the resultant $\omega_{2n}$'s is very small at the beginning and will increase significantly when $r_0$ is big enough. This kind of
phenomena are consistent with our Figs. 2, 3 and 4 in \cite{Chao2021}.

\subsection{Test the isospectrality}

As we have seen in Tables \ref{table1} - \ref{table3}, there is almost no deviation on $\omega$'s between the axial- and polar-perturbation cases up to our numerical error. Therefore, we have confirmed the isospectrality for those scenarios. Of course, in there we were using the values for model-dependent constants (cf. $r_0$ and $\rho_0$) based on the current observations. Nonetheless, as mentioned
previously, $\rho_0$ and $r_0$ are basically free parameters. It is worth testing if we can break the isospectrality with large enough $\rho_0$ and $r_0$.

\begin{figure}[htb]
	\includegraphics[width=\columnwidth]{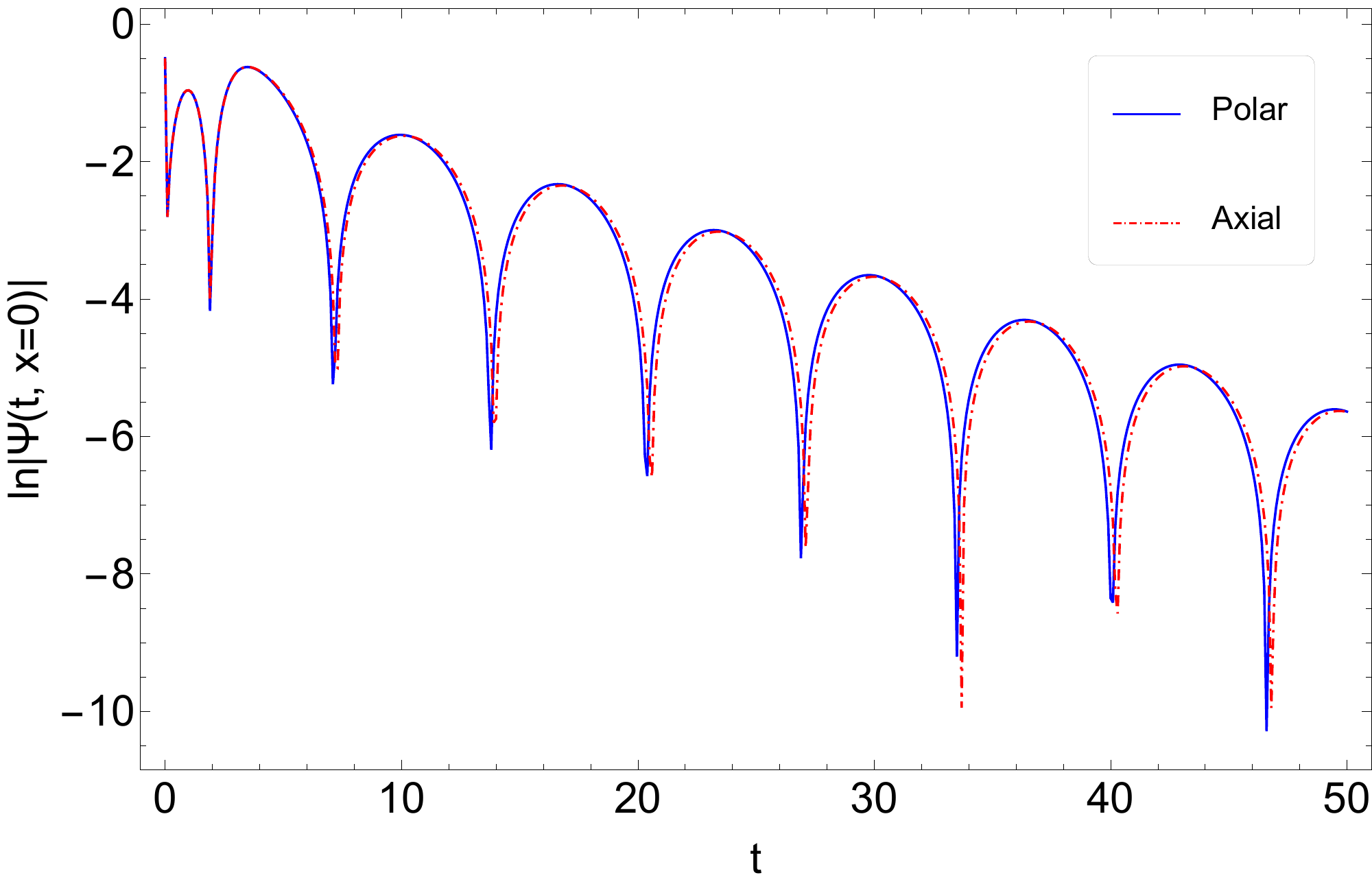}
	\includegraphics[width=\columnwidth]{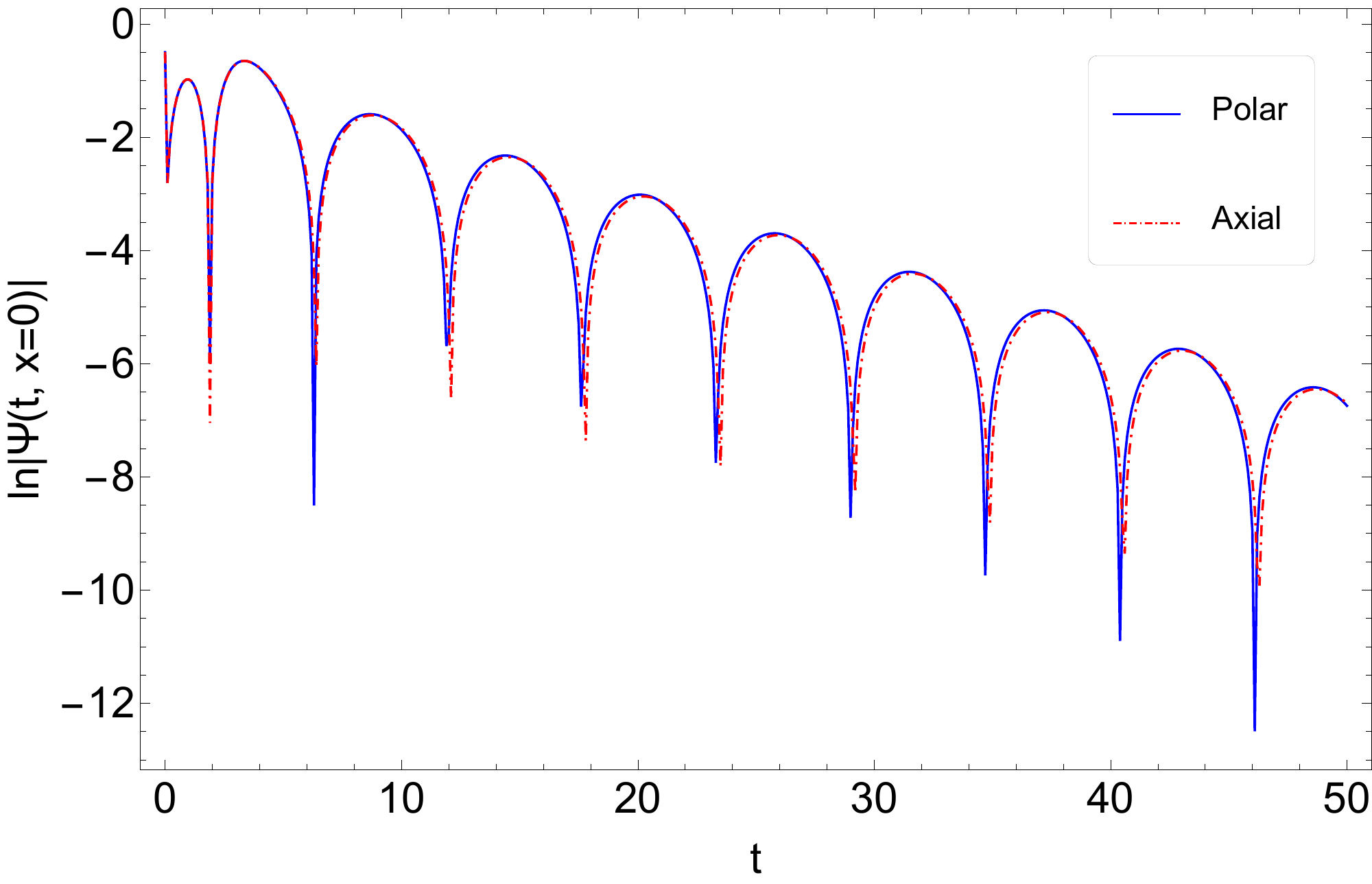}
	\caption{ The temporal evolution of $\Psi(t, x=0)$ [cf. \eqref{master1} and \cite{Chao2021}'s (3.9)] under the polar$\setminus$axial perturbations for the Case 3 in Table. \ref{table0} by setting $l=2$. For the upper panel we have $r_0=15.7 \text{kpc}$ and $\rho_0=10^{13} \text{M}_{\odot}/{\rm kpc}^{3}$. For the lower panel we have $r_0=7000 \text{kpc}$ and $\rho_0=3.43\times10^{7} \text{M}_{\odot}/{\rm kpc}^{3}$. Here, the blue solid lines and red dash-dotted lines represent the results from the polar and axial perturbations, respectively. Note that here we are using the unit system so that $c=G_N={r_{MH}^{\text{Sch}}}=1$.}
	\label{plot3}
\end{figure}

In this subsection, we test the influence of $\rho_0$ and $r_0$ on the isospectrality. For this part, we continue to work with Case 3. By using \eqref{master1}, \cite{Chao2021}'s (3.9) and \eqref{FDM1}, we could carry out the calculations. The resultant $\ln |\Psi(t, x=0)|$ for the polar and axial cases are compared in Fig. \ref{plot3}. In there we set  $l=2$. In addition, for the upper panel we set $r_0=15.7 \text{kpc}$ and $\rho_0=10^{13} \text{M}_{\odot}/{\rm kpc}^{3}$, while for the lower panel we set $r_0=7000 \text{kpc}$ and $\rho_0=3.43\times10^{7} \text{M}_{\odot}/{\rm kpc}^{3}$. Here we have made the $\rho_0$  and $r_0$ big enough for upper and lower panel respectively to break the  isospectrality (suppose we can). However, it turns out that we can barely observe deviations between these two cases (axial and polar) since the two curves in both the upper and lower panels of Fig. \ref{plot3} are almost overlapped. Therefore, we conclude that the isospectrality is preserved. Besides, just like we anticipated, for Case 1 and Case 2 we can also observe similar phenomena. For simplicity, we omit the redundant details here.

\section{Implication in gravitational wave detection}
\renewcommand{\theequation}{5.\arabic{equation}} \setcounter{equation}{0}

Once we are able to calculate QNMs from the master equations that govern the axial and polar perturbations of the Schwarzschild-like BHs with DM, our purpose here is to see how the shifts on QNM frequencies affect the GWs from the ringdown stages of coalescences. We are mostly  interested in the shifts in the Schwarschild QNM frequencies induced by the presence of DM halos sorrounding a supermassive BH, characterized by
\bqn
\omega = \omega^{\rm Sch} + \delta \omega,
\eqn
where $\omega^{\rm Sch}$ denotes the QNM frequencies of the Schwarzschild black hole and $\delta \omega$ denotes the corrections from the dark matter halo. Here we would like to mention that a remarkable result in GR is the isospectrality of QNM of the Schwarzschild and Kerr BHs. Normally, when the Schwarzschild geometry is perturbed by the DM halo, isospectrality will in general be broken. However, as we have shown in the above section, the isospectrality is still satisfied within the tolerance of the numerical errors. In this case we will treat  the polar and axial perturbations as isospectral.

Now we need to construct the corresponding GW waveform. The GW emitted during the ringdown stage can be expressed as a linear combination of dampled sinusoids,
\bqn
h_{+} + i h_{\times} =\frac{M_z}{D_{\rm L}} \sum_{lmn} {\cal A}_{lmn} e^{i(f_{lmn} t + \phi_{lmn}) } e^{- t/\tau_{lmn}} S_{lmn},\nb\\
\eqn
where $M_z$ is the red-shifted mass of the BH, $D_{\rm L}$ is the luminosity distance to the source, ${\cal A}_{lmn}$ is the mode amplitude, $\phi_{lmn}$ is the phase coefficient,  and $S_{lmn}$ is the (complex) spin-weighted spheroidal harmonics of spin weight 2, which depend on the polar and azimuthal angles. The GW frequency satisfies $2 \pi f_{lmn} = \text{Re}(\omega_{lmn})$, with the right-hand side being the real part of the QNM frequency for the $(l, m, n)$ mode, while the damping time $\tau_{lmn}$ is related to the imaginary part of QNM frequency via $\tau_{lmn} =  -1/\text{Im}(\omega_{lmn})$ \cite{Berti2006}. To illustrate the effects of DM, we can express the frequency $f_{lmn}$ and damping time $\tau_{lmn}$ in terms of (small) deviations to the corresponding Schwarzschild values,
\bqn
\lb{deltaf}
f_{lmn} = f_{lmn}^{\rm Sch} (1+ \delta f_{lmn}), \\
\tau_{lmn} = \tau_{lmn}^{\rm Sch} (1+ \delta \tau_{lmn}),
\eqn
where $ f_{lmn}^{\rm Sch} $ and $ \tau_{lmn}^{\rm Sch} $ are the QNM frequency and damping time of the Schwarzschild BH, and $ \delta f_{lmn}$ and $\delta \tau_{lmn}$ represent the deviations from the Schwarzschild case due to presence of DM halos.
$ f_{lmn}^{\rm Sch} $ and $ \tau_{lmn}^{\rm Sch} $ only depend on the mass of a BH and the fundamental mode with $n=0, l=2$, $m = 0$  has frequency and damping time (in SI unit)  \cite{Berti2006}
\bqn
\lb{f220}
f_{200}^{\rm Sch}  \simeq 0.012 {\rm Hz}  \frac{10^6 M_{\odot}}{M}, \\
\lb{tau220}
\tau_{200}^{\rm Sch}  \simeq 55.4 {\rm s}  \frac{10^6 M_{\odot}}{M}.
\eqn

\begin{figure}[htb]
	\includegraphics[width=\columnwidth]{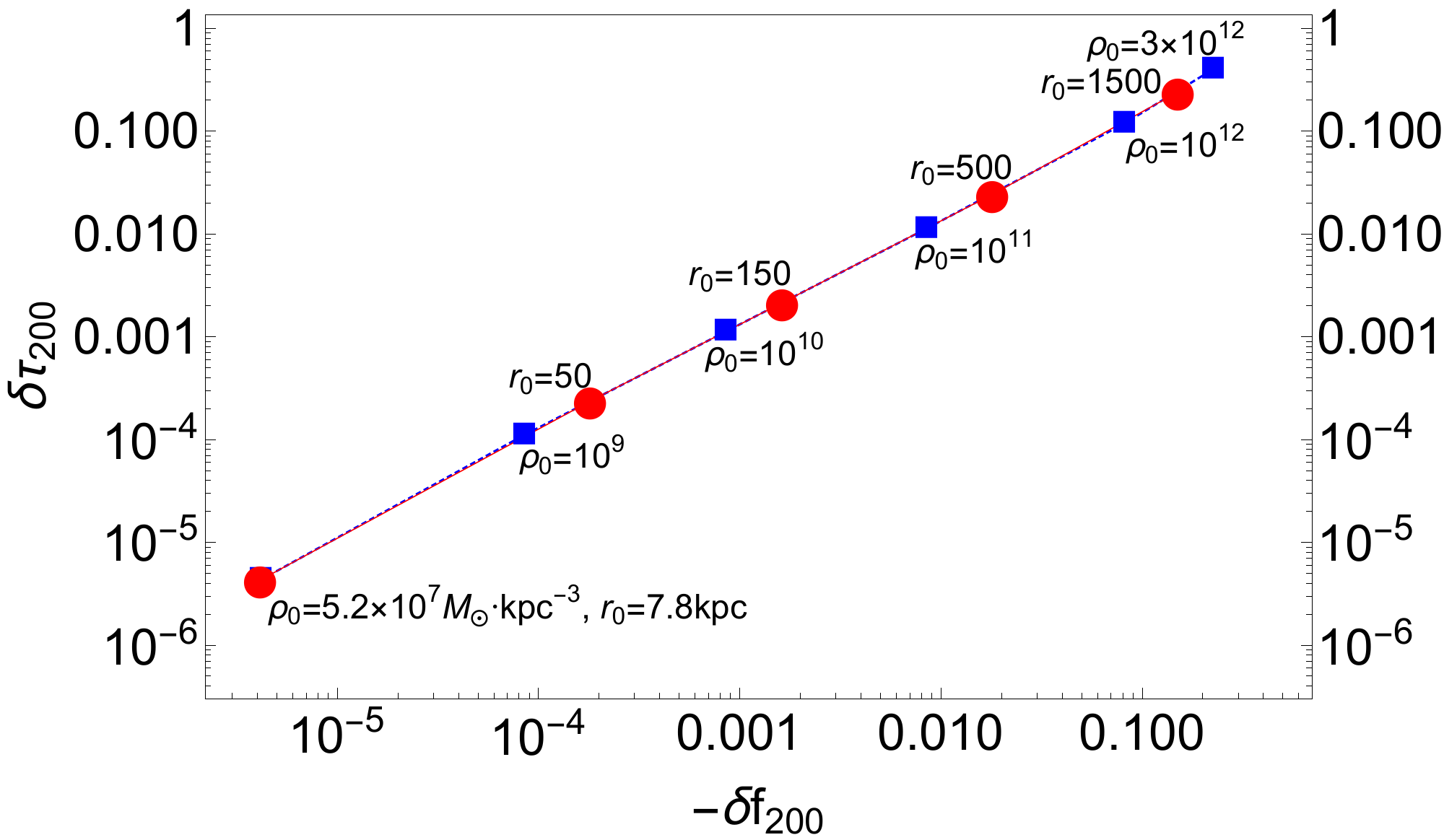}
	\includegraphics[width=\columnwidth]{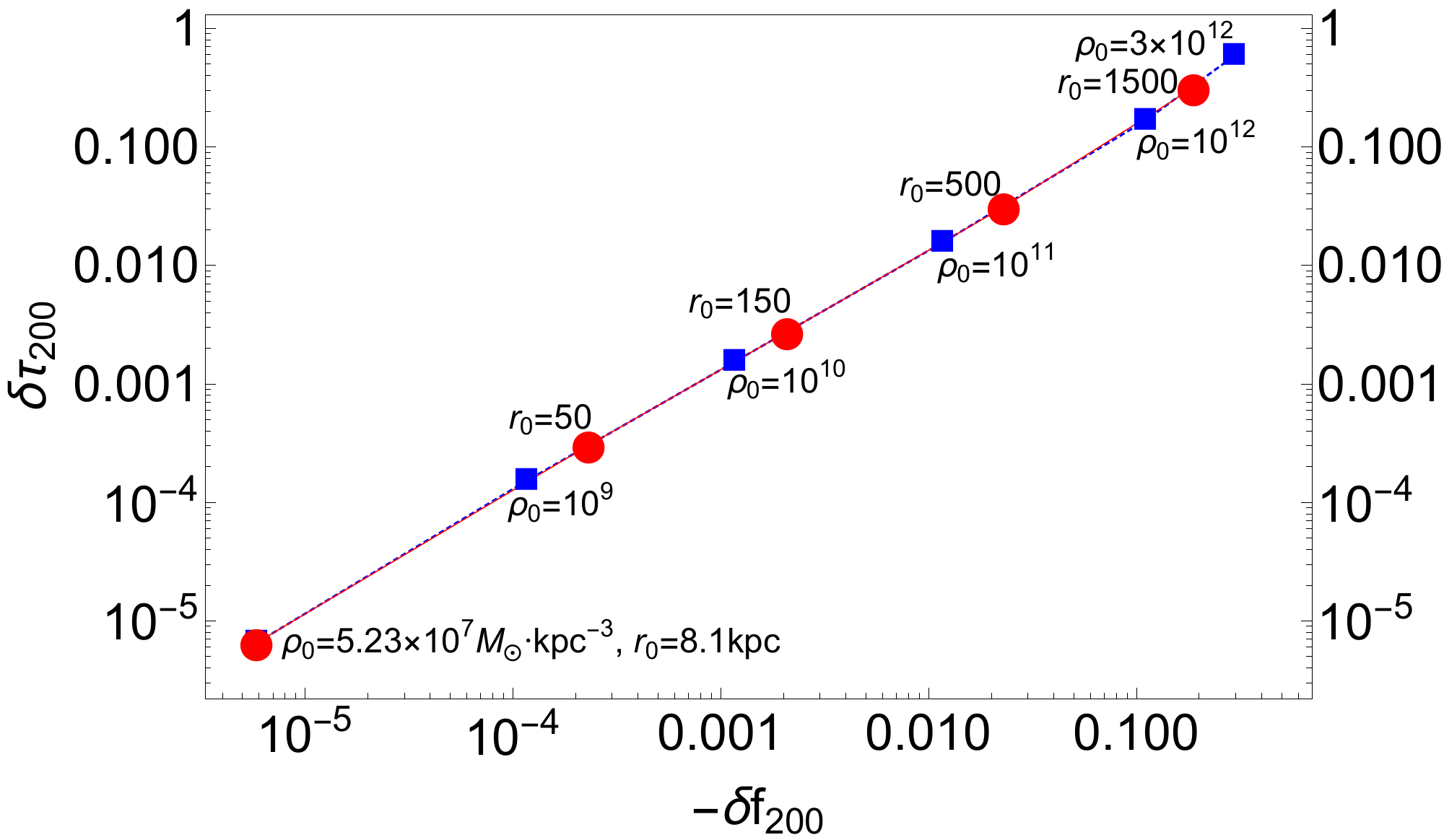}
	\includegraphics[width=\columnwidth]{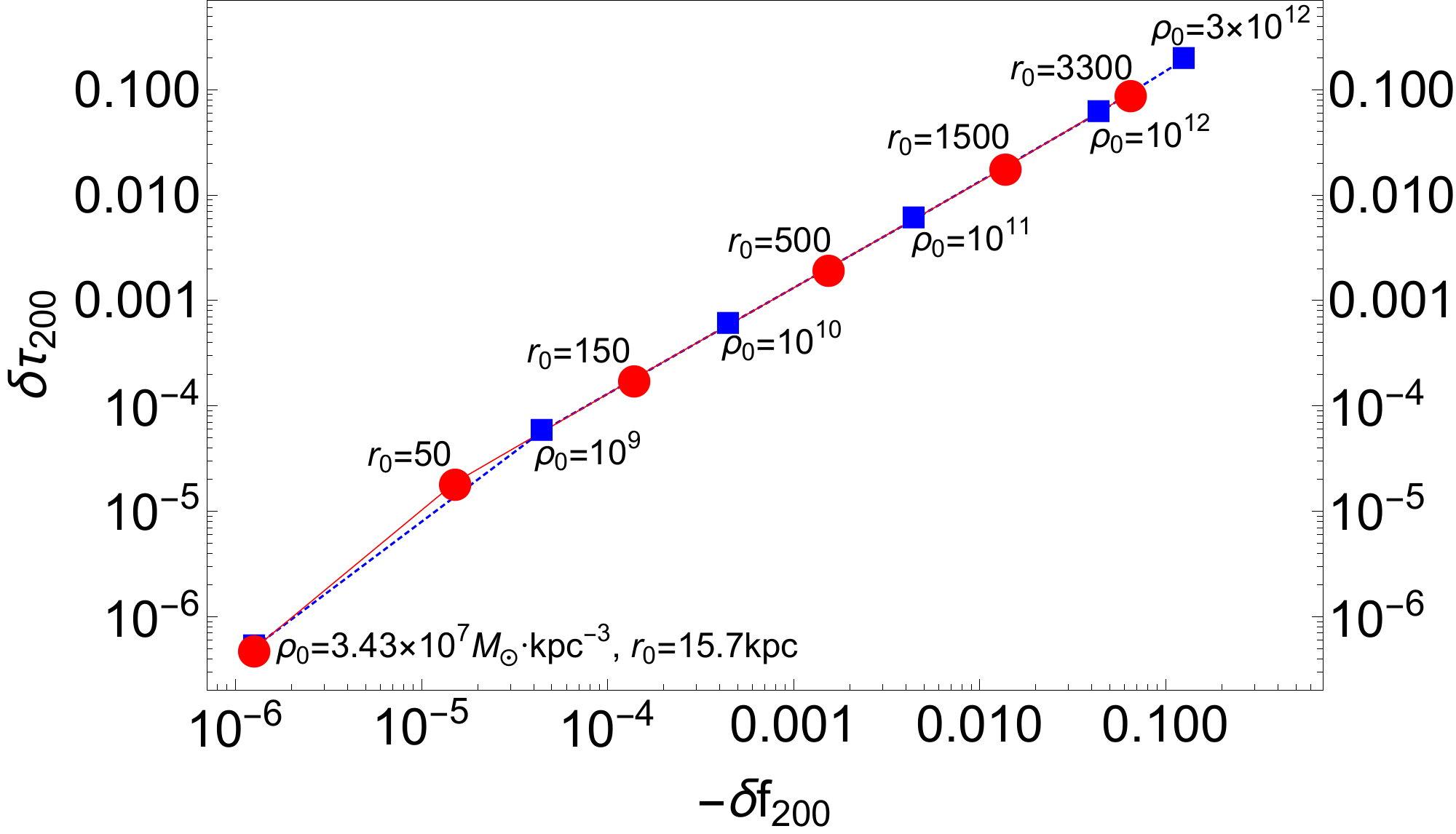}
	\caption{ The deviations of QNM frequency $ \delta f_{200}$ and damping time $\delta \tau_{200}$ from those of a  Schwarzschild BH by considering the  Sgr $\text{A}^\ast$ BH for different values of $\rho_0$ and $r_0$. Upper panel: Case 1; middle panel: Case 2; and bottom panel: Case 3 (cf. Table \ref{table0}). The blue dashed lines represent the results for fixing $r_0$ (They are also marked by little blue boxes), while the red solid lines represent the results for fixing $\rho_0$ (They are also marked by red dots).  Note that here we are using the unit system so that $c=G_N={r_{MH}^{\text{Sch}}}=1$.}
	\label{deltaw}
\end{figure}

The $ \delta f_{lmn}$ and $\delta \tau_{lmn}$ are functions of the mass of a BH, viz., $M$, and the parameters $\rho_0$ and $r_0$ are given by different DM profiles. Now we expect to see how $ \delta f_{lmn}$ and $\delta \tau_{lmn}$  depend on $\rho_0$ and $r_0$, and whether they are detectable with future space-based GW detectors. We investigate the Sgr $\text{A}^\ast$ BH and illustrate the behaviors of $ \delta f_{200}$ and $\delta \tau_{200}$ for different values of $\rho_0$ and $r_0$ in Fig.~\ref{deltaw} by considering the three different cases (cf. Table \ref{table0}). The calculations are, again, carried out by using the matrix method. From this figure we observe greater $-\delta f_{200}$ and $\delta \tau_{200}$ with increasing $r_0$ or $\rho_0$. This indicates that a denser DM distribution in the central region of a galaxy near a BH leads to lower GW frequency and longer damping time for GWs during the ringdown stage. It is also evident that $-\delta f_{200}$ and $\delta \tau_{200}$ can be as large as $10^{-1}$ for certain values of $\rho_0$ and $r_0$. Considering {the designed resolution} of space-based detectors, such as LISA, TianQin, and Taiji, these effects may be detectable \cite{Shi2019}. This may provide an approach to probe the matter distribution in the central region of a galaxy.

Note that in the above analysis, we were considering QNMs of Schwarzschild-like BHs, i.e., we ignored the rotation of a BH for simplicity. Including the effects of rotations of a BH is not an easy task since the exact rotating solution in the DM halo is still lacking. It is also quite difficult to construct the master equations for gravitational perturbations of a rotating BH beyond Kerr. Such effects may be considered in our future works.

\section{Conclusion and Discussions}
\renewcommand{\theequation}{5.\arabic{equation}} \setcounter{equation}{0}

In this paper, we investigate the calculations of QNMs and focus on supermassive BHs in the central region of a galaxy surrounded by a DM halo, which are described by Schwarzschild-like spacetimes. With three different DM models, their metrics are given by \eqref{URC_Gr}, \eqref{Gr2} and \eqref{Gr3}, respectively. Each model is mainly measured by two model-dependent parameters, viz., $\rho_0$ and $r_0$. For the two specific BHs that we studied, i.e., the Sgr A$^*$ BH (located at the cener of Milky Way) as well as the M87 central BH, the corresponding values of $\rho_0$ and $r_0$ are summarized in Table \ref{table0}. In addition to the Schwarzschild case, the other three non-Schwarzschild cases are referred as Case 1, Case 2 and Case 3 in Table \ref{table0}.

To calculate QNMs, a master equation is derived under the background \eqref{backg2} and the perturbation \eqref{hab}, where the RW gauge \cite{Thomp2017} is what we adopt. Notice that, in \cite{Chao2021} the axial (odd-parity) perturbation has been studied, and here we focus on the polar (even-parity) sector. By using the Einstein's field equations \cite{CarrollB} and following \cite{Wentao2021}, our result of master equation for the polar sector is given by \eqref{master1}, with an effective potential given by \eqref{Veff}. Since the resultant QNM frequency, viz., $\omega$, is directly related to the patterns of an effective potential, we want to first take a loot to that. For this purpose, $V_{\text{eff}}$'s are plotted in Fig. \ref{plot1} for the four cases listed  in Table \ref{table0} (including the Schwarzschild case).

Since the curves in Fig. \ref{plot1} are almost overlapped, we anticipate that the resultant $\omega$'s [cf. \eqref{master2}] will be quite similar. This is confirmed by using the results shown in Tables \ref{table1} - \ref{table3}, in which we provide the calculated $\omega$'s by setting $l=2, 3, 4, 5$ and $n=0, 1$ (Recall that $\omega$'s could be calculated for different modes, distinguished by $l$, $m$ and $n$, and we can denote them by $\omega_{lmn}$ or $\omega_{ln}$, since we have set $m=0$ due to the fact that we are dealing with spherically symmetric spacetimes) for the polar, axial cases. To find these $\omega$'s, two techniques are utilized, namely, the sixth-order WKB method [cf. \eqref{WKB1}] as well as the matrix method [cf. \eqref{detK}]. By using them we can guarantee the accuracy of our results.

From  Tables \ref{table1} - \ref{table3}, we notice that, for a considered $\omega_{ln}$, the discrepancy between each case (axial, polar and Schwarzschild) occurs at the 4th digit or after that. For most of the  $\omega_{ln}$'s appearing in Tables \ref{table1} - \ref{table3}, such a discrepancy is quite negligible, so that we conclude that  the various cases listed in Table \ref{table0} won't make too much difference. In addition, by comparing the results from the axial and polar perturbations, we can also confirm the  isospectrality \cite{Berti2009} for the three DM models and the parameters considered in Table \ref{table0}.

At the same time, since the values of $\rho_0$ and $r_0$ may change from galaxy to galaxy, we also investigate the impacts of them on QNMs. {Indeed, according to the current observations to the Milky Way and M87 galaxies, the resultant $\rho_0$ and $r_0$ will lead to quite negligible deviations from the Schwarzschild case, as we have seen from  Tables \ref{table1} - \ref{table3}. Nonetheless, it's hard to tell what kind of parameters we will obtain for other galaxies in the universe. Thus, it is worth checking what will happen when $\rho_0$ and $r_0$ are changing freely. More importantly, in the context of constraints on BH environments, these parameters are basically free \cite{Cardoso:2021wlq}. It's not necessary to assume that they will preserve the similar magnitudes in all the occasions. A more reasonable way is to consider $\rho_0$ and $r_0$ on a wider range. This fact stimulates our interests on how these parameters will influence QNMs when they are changing freely.} Besides, since the $l=2$ mode is in general the dominate one \cite{Mag18}, for this part we focus on the $l=2$ mode only.

Basically, we study the influence of one of the two parameters $\rho_0$ and $r_0$  on QNMs by fixing the other. Treating Case 3 (cf. Table \ref{table0}) as our example, the final results are exhibited in Fig. \ref{plot2}. In obtaining these results, to illustrate more clearly how different cases with varying  $\rho_0$ and $r_0$ deviate from each other, a new technique, viz., the FDM, is applied [cf. \eqref{FDM1}]. By using the FDM, our master variable could be solved in the time domain, and we are able to obtain $\Psi (t, x=0)$ [cf. \eqref{master1}], which carries information of all the existing $\omega_{2n}$'s. From Fig. \ref{plot2} we learn that, different  $\rho_0$'s and $r_0$'s will result in almost identical results when they are small. Nevertheless, once $\rho_0$'s or $r_0$'s are large enough, the resultant $\Psi (t, x=0)$ (so that $\omega_{2n}$'s) will be sensitive to the values of $\rho_0$ and $r_0$. In fact, this phenomenon is consistent with what we have observed in \cite{Chao2021}.

On the other hand, to further confirm the isospectrality, we apply the FDM once again to Case 3 of Table \ref{table0} by setting $l=2$ and changing $\rho_0$ and $r_0$ to much larger values. Although $\rho_0$ and $r_0$ are adjusted to very large quantities, we can barely observe deviations between the axial and polar cases, as seen from Fig. \ref{plot3}. Therefore, by combining the observations from Tables \ref{table1} - \ref{table3} and  Fig. \ref{plot3}, we conclude that the isospectrality is preserved in the three DM models we consider here (cf. Sec. II) \footnote{Recall that here we have ignored the perturbation of the DM. It is still an open question whether the isospectrality could be preserved once the perturbation of DM presents.}.

Finally, we want to investigate in detail how QNMs deviate from that of the Schwarzschild case  by changing the model-dependent parameters  $\rho_0$ and $r_0$. For this purpose, we characterize such deviations by $\delta f_{lmn}$ and $\delta \tau_{lmn}$ [cf. \eqref{deltaf}]. Here, {we focus on the $l=2$, $n=0$ and $m=0$ mode,} since this is one of the dominate ones.
As an example, we investigate the Sgr $\text{A}^\ast$ BH and illustrate the behaviors of $ \delta f_{200}$ and $\delta \tau_{200}$ for different values of $\rho_0$ and $r_0$ in Fig.~\ref{deltaw} by considering the three different DM models (cf. Table \ref{table0}).
From this figure we observe greater $-\delta f_{200}$ and $\delta \tau_{200}$ with increasing $r_0$ and $\rho_0$. This indicates that a denser DM distribution in the central region of a galaxy near a BH leads to lower GW frequency and longer damping time for GWs during the ringdown stage. It is also evident that $-\delta f_{200}$ and $\delta \tau_{200}$ can be as large as $10^{-1}$ for certain values of $\rho_0$ and $r_0$.
 By considering the designed resolution of LISA-like detectors \cite{Shi2019}, one may expect such a large deviation can be found in reality once a galaxy with suitable $\rho_0$ and $r_0$ is observed someday. By matching with the results here, this kind of observations will either confirm our current understanding to DM or help us put constraints on the current DM models.

Our work here can be extended in several directions. First of all, here we only consider three different DM profiles. It is interesting to extend the current work to other DM profiles. For instance, BHs surrounded by superfluid DM and baryonic matter \cite{Kimet2020}. In addition, as has been mentioned, the higher-order effect of potential of DM is ignored in the current study  \cite{xu_JCAP}, so that we can assume $G(r)=F(r)$ [cf. \eqref{metric_dark}]. A natural desire is to extend our work to more general cases, for which we can relax such an assumption. Inspired by \cite{Zhaoyi2020, Cardoso2022, Konoplya2022, Kimet2022}, we may run a systematic study in this direction in our next step. On the other hand, since astrophysical BHs in general have non-zero angular momentum, it is also our plan to extend our work to rotating BHs. Finally, we may test the effects of DM halos on various modified theories of gravity.

\section*{Acknowledgements}

We appreciate the helpful discussions with Hao-Jie Lin. This work is supported by the National Key Research and Development Program of China Grant No.2020YFC2201503, the Zhejiang Provincial Natural Science Foundation of China under Grant No. LR21A050001 and LY20A050002, the National Natural Science Foundation of China under Grant No. 11675143, No. 11975203 and No. 11705053, and the Fundamental Research Funds for the Provincial Universities of Zhejiang in China under Grant No. RF-A2019015.

\end{document}